\documentclass[twocolumn]{aastex63}

\usepackage{amsmath, bm}
\usepackage{color}
\usepackage{textcomp}
\usepackage{url}
\usepackage{graphicx}
\usepackage{subfigure}
\usepackage{listings}
\usepackage{hyperref}
\usepackage{enumitem}
\usepackage{longtable}
\usepackage{lineno}

\begin{document}

\newcommand{\rvshift}{$\sim 0.6$\,\kms}
\newcommand{\effectorbit}{$\sim 1.5$\,Myr}
\newcommand{\effectrvshift}{$\sim 2$\,Myr}
\newcommand{\effectcontamination}{$\sim 3$\,Myr}
\newcommand{\effecterror}{$\sim 0.6$\,Myr}
\newcommand{\ageuncedr}{$17.8 \pm 1.6$\,Myr}
\newcommand{\agecoredr}{$18.4 \pm 1.6$\,Myr}
\newcommand{\ageunc}{$19.8 \pm 2.5$\,Myr}
\newcommand{\agecor}{$20.4 \pm 2.5$\,Myr}
\newcommand{\agecrundall}{$17.8 \pm 1.2$\,Myr}
\newcommand{\agemiretroigtrace}{$18.5^{+2.0}_{-2.4}$\,Myr}
\newcommand{\agemiretroigdet}{$17.5^{+3.5}_{-2.9}$\,Myr}
\newcommand{\inputsample}{$76$}
\newcommand{\coresample}{$25$}
\newcommand{\extendedsample}{$51$}
\newcommand{\nstarsinitfirst}{37}
\newcommand{\nstarsinit}{33}
\newcommand{\nmodelmembers}{25}
\newcommand{\nwhitedwarfs}{12}
\newcommand{\ngaiasearch}{503}
\newcommand{\ninternalcomovers}{52}
\newcommand{\ninternalsystems}{26}
\newcommand{\nalonecomovers}{2}
\newcommand{\npartialcomovers}{21}
\newcommand{\nvisualcomovers}{12}
\newcommand{\npartialsystems}{15}
\newcommand{\slfrac}[2]{\left.#1\middle/#2\right.}

\newcommand{\bpmg}{$\beta$PMG}
\newcommand{\bpic}{$\beta$~Pictoris}
\newcommand{\bpicfull}{\bpic\ moving group}
\newcommand{\bpictext}{\texorpdfstring{\bpic}{Beta~Pictoris} Moving Group}
\newcommand{\xu}{$X--U$}
\newcommand{\yv}{$Y--V$}
\newcommand{\zw}{$Z--W$}
\newcommand{\xez}{$\xi^\prime\eta^\prime\zeta^\prime$}
\newcommand{\majasso}{$\mu$~TAU}
\newcommand{\hip}{\emph{Hipparcos}}
\newcommand{\gaia}{\emph{Gaia}}
\newcommand{\gaiagr}{$G - G_{\rm RP}$}
\newcommand{\gaiar}{$G_{\rm RP}$}
\newcommand{\gaiab}{$G_{\rm BP}$}
\newcommand{\gaiadr}{\gaia~DR3}
\newcommand{\gaiaedr}{\gaia~EDR3}
\newcommand{\gaiadrtwo}{\gaia~DR2}
\newcommand{\kms}{\hbox{km\,s$^{-1}$}}
\newcommand{\mjup}{$M_{\mathrm{Jup}}$}
\newcommand{\rjup}{$R_{\mathrm{Jup}}$}
\newcommand{\msol}{$M_{\odot}$}
\newcommand{\rsol}{$R_{\odot}$}
\newcommand{\masyr}{$\mathrm{mas}\,\mathrm{yr}^{-1}$}
\newcommand{\teff}{$T_{\rm eff}$}
\newcommand{\lbol}{$\log{L_*/L_\odot}$}
\newcommand{\besancon}{Besan\c con}
\newcommand{\banyanE}{BANYAN~$\Sigma$}

\newcolumntype{z}{>{\raggedright\arraybackslash}p{15.8cm}}

\accepted{January 18th 2023}
\submitjournal{ApJ}
\title{Addressing Systematics in the Traceback Age of the \bpictext}
\shorttitle{Systematics in the Traceback Age of the \bpmg}
\shortauthors{Couture et al.}

\author[0000-0003-2604-3255]{Dominic Couture}
\affiliation{Plan\'etarium Rio Tinto Alcan, Espace pour la Vie, 4801 av. Pierre-de Coubertin, Montr\'eal, Qu\'ebec, Canada}
\affiliation{Trottier Institute for Research on Exoplanets, Universit\'e de Montr\'eal, D\'epartement de Physique, C.P.~6128 Succ. Centre-ville, Montr\'eal, QC H3C~3J7, Canada}
\affiliation{D\'epartement de Physique, Universit\'e de Montr\'eal, C.P.~6128 Succ. Centre-ville, Montr\'eal, QC H3C~3J7, Canada}
\email{couture@astro.umontreal.ca}

\author[0000-0002-2592-9612]{Jonathan Gagn\'e}
\affiliation{Plan\'etarium Rio Tinto Alcan, Espace pour la Vie, 4801 av. Pierre-de Coubertin, Montr\'eal, Qu\'ebec, Canada}
\affiliation{Trottier Institute for Research on Exoplanets, Universit\'e de Montr\'eal, D\'epartement de Physique, C.P.~6128 Succ. Centre-ville, Montr\'eal, QC H3C~3J7, Canada}
\email{gagne@astro.umontreal.ca}

\author[0000-0001-5485-4675]{René Doyon}
\affiliation{Trottier Institute for Research on Exoplanets, Universit\'e de Montr\'eal, D\'epartement de Physique, C.P.~6128 Succ. Centre-ville, Montr\'eal, QC H3C~3J7, Canada}
\affiliation{D\'epartement de Physique, Universit\'e de Montr\'eal, C.P.~6128 Succ. Centre-ville, Montr\'eal, QC H3C~3J7, Canada}
\email{doyon@astro.umontreal.ca}

\begin{abstract}
 We characterize the impact of several sources of systematic errors on the computation of the traceback age of the \bpicfull\ (\bpmg). We find that uncorrected gravitational redshift and convective blueshift bias absolute radial velocity measurements by \rvshift, which leads to erroneously younger traceback ages by \effectrvshift. Random errors on parallax, proper motion, and radial velocity measurements lead to an additional bias of \effecterror\ on traceback ages. Contamination of astrometric and kinematic data by kinematic outliers and unresolved multiple systems in the full input sample of \inputsample\ members and candidates of \bpmg\ also erroneously lowers traceback ages by \effectcontamination. We apply our new numerical traceback analysis tool to a core sample of \coresample\ 
 carefully vetted members of \bpmg\ using \gaia\ Data Release~3 (DR3) data products and other kinematic surveys. Our method yields a corrected age of \agecor, bridging the gap between kinematic ages ($11-19$\,Myr) and other age-dating methods, such as isochrones and lithium depletion boundary ($20-26$\,Myr). We explore several association size metrics that can track the spatial extent of \bpmg\ over time, and we determine that minimizing the variance along the heliocentric curvilinear coordinate $\xi^\prime$ (i.e., toward the Galactic Center) offers the least random and systematic errors, due to the wider $UVW$ space velocity dispersion of members of \bpmg\ along the $U$-axis, which tends to maximize the spatial growth of the association along the $\xi^\prime$-axis over time.
\end{abstract}

\section{Introduction}
\label{sec:introduction}

Nearby young associations (NYAs) are sparse, coeval, gravitationally unbound stellar populations located in the solar neighborhood. Formed within a few million years of each other from the collapse of a single molecular cloud or cloud complex, members of these kinematic associations have similar Galactic positions and space velocities, and they share the same age and chemical composition due to their common formation history. However, their members span relatively large projected angular areas on the sky, as a result of the proximity and low spatial density of NYAs, which has made identification of NYAs challenging, especially prior to the \gaia\ mission \citep{2016AA...595A...1G}.

Such populations of nearby, age-calibrated stars are ideal laboratories to study the last stages of stellar and exoplanetary formation, and they provide strategic locations to search for isolated planetary-mass objects and for the direct imaging of giant exoplanets, thanks to the more favorable contrast between the host star and its companion, which remain relatively bright in the near-infrared (NIR) at such young ages. This contrast is even more advantageous for low-mass, late-type stars, which make up the majority of NYAs.

For up to a few hundred million years after their formation, members of an NYA retain similar kinematics, until gravitational perturbations cause their Galactic orbits to become randomized enough that they become indistinguishable from unrelated field stars. Using the full 6D kinematics of their members (i.e., their $XYZ$ Galactic positions and $UVW$ space velocities, where $U = dX/dt$, $V = dY/dt$, and $W = dZ/dt$), it is possible to compute backward Galactic orbits and trace members' trajectories back to the epoch when the NYA's spatial extent was minimal, which is assumed to coincide with the epoch of stellar formation. Thus, traceback analysis can provide a kinematic age estimate for members of an NYA, independent of stellar evolution models, unlike the more usual isochrone or lithium depletion boundary (LDB) methods.

One of the greatest difficulties in tracing back stellar trajectories is the need for precise astrometric and kinematic measurements to compute the full 6D kinematics of members of an NYA. The \gaia\ Data Release~3 (DR3) data products provide an unmatched sample of precise parallax and proper motion measurements \citep{2022arXiv220800211G}. However, despite the increase of available measurements over the \gaia\ Early Data Release~3 \citep{2021AA...649A...1G}, absolute radial velocity measurements remain relatively inaccurate and represent, by far, the largest contribution to the total error in 6D kinematics during backward Galactic orbit integration, especially given how errors on Galactic position grow as stars are projected further back in time.

Jointly discovered by \cite{1999ApJ...520L.123B} and \cite{2001ApJ...562L..87Z}, using data from the \hip\ catalog, the \bpicfull\ (\bpmg) is one of the youngest and nearest known NYAs. Over 40 stars, located at an average distance $\sim 35$\,pc, with a distance range of $9-72$\,pc, with computed full 6D kinematics and visible signs of youth consistent with membership in \bpmg, are considered to be \emph{bona fide} members of the association, and more than a hundred additional candidate members are known, for which kinematic measurements for final membership determination are needed \citep{2006AA...460..695T, 2010AJ....140..119S,
2011MNRAS.411..117K, 2012AJ....143...80S, 2014ApJ...788...81M,
2014AJ....147...85R, 2016MNRAS.455.3345B, 2017AJ....153...95R, 2018ApJ...860...43G, 2020AA...642A.179M}.

The youth and proximity of \bpmg\ make its members ideal targets for the search and characterization of exoplanets through direct imaging. For instance, its eponymous star, \bpic, has two known giant exoplanets, a debris disk, and exocomets \citep{2009AA...493L..21L, 2012AA...542A..41C, 2014Natur.514..462K, 2019NatAs...3.1135L}. PSO~J318.5338--22.8603, a $6.5^{+1.3}_{-1.0}$\,\mjup\ free-floating planetary-mass object of spectral type L7, was discovered using data from the Pan-STARRS survey \citep{2013ApJ...777L..20L}. With an $XYZ$ Galactic position and an $UVW$ space velocity compatible with membership in \bpmg, its age is the same as the rest of the association. 

However, traceback age estimates for \bpmg\ and other NYAs are inconsistent with other age-dating methods. Age estimates for \bpmg\ using the isochrones and LDB methods ($20-26$\,Myr; \citealp{2008ApJ...689.1127M, 2010ApJ...711..303Y, 2014ApJ...792...37M, 2014MNRAS.445.2169M, 2014MNRAS.438L..11B, 2015MNRAS.454..593B, 2022AA...664A..70G}) are significantly older than most traceback age estimates ($11-13$\,Myr; \citealp{2002ApJ...575L..75O, 2003ApJ...599..342S, 2004ApJ...609..243O, 2018AA...615A..51M}), although more recent kinematic approaches ($17-19$\,Myr; \citealp{2019MNRAS.489.3625C, 2020AA...642A.179M}) have addressed specific and distinct sources of bias, and have allowed the gap between isochrones and LDB methods to be bridged. \cite{2019MNRAS.489.3625C} used a forward modeling approach, which circumvents the bias on traceback ages due to measurement errors (see Section~\ref{sec:sensitivity_kinematic_errors}) that arises when computing backward Galactic orbits for individual members, whereas \cite{2020AA...642A.179M} used a more traditional traceback approach but minimized sample contamination through a rigorous selection process and used robust association size metrics to track the spatial extent of \bpmg\ over time.

In this work, we aim to correct for various sources of systematic errors in the computation of the traceback age of \bpmg, and to determine whether this can further reconcile the tension between kinematic ages and other age-dating methods. Our approach uses a clean sample of \emph{bona fide} members of \bpmg, free of any kinematic outlier or unresolved multiple system (see Section~\ref{sec:sample_selection}), along with data from \gaiadr\ data products and other radial velocity surveys, as the basis to compute backward Galactic orbits using their full 6D kinematics. We account for the bias on traceback age estimates due to measurement errors in astrometric and kinematic data (see Section~\ref{sec:sensitivity_kinematic_errors}) and two separate biases on radial velocity measurements, gravitational redshift and convective blueshift (see sections~\ref{sec:grav_redshift}~and~\ref{sec:conv_blueshift}), all of which tend to artificially push the epoch of minimal association size closer to the current-day epoch. We also test several association size metrics to evaluate the spatial extent of NYAs over time, in order to determine which provide the most accurate and reliable age estimates (see Section~\ref{sec:metrics}).

This study is structured as follows. First, in Section~\ref{sec:sample_selection}, we describe how we selected a clean, vetted sample of \emph{bona fide} members of \bpmg\ and corrected for biases on radial velocity measurements. In Section~\ref{sec:traceback_method}, we describe the numerical method used to derive a kinematic age estimate of NYAs, and we apply it to simulated samples of stars in order to assess the precision and bias of every association size metric. In Section~\ref{sec:results_discussion}, we apply our method to our sample of members of \bpmg, and we compare our results to previous age estimates for \bpmg. Finally, we conclude our analysis in Section~\ref{sec:conclusions}.

\section{Sample selection}
\label{sec:sample_selection}

When assembling a sample with the aim to perform traceback analysis, candidates must be carefully vetted in order to minimize contamination by unrelated older stars that show a kinematic compatible with the NYA. Indeed, our simulations show that traceback ages are not only less precise with the addition of kinematic outliers, they are also biased toward younger ages because the positions of kinematic outliers do not converge at the epoch of stellar formation along with actual members of the NYA (see Section~\ref{sec:simulated_samples}).

Therefore, in order to mitigate this issue, we first assembled a sample of candidate members of \bpmg\ from the literature \citep{2001ApJ...562L..87Z, 2013ApJ...762...88M, 2014AJ....147...85R, 2020AA...642A.179M} with available full 6D kinematics. Candidates were also vetted by the Bayesian analysis tool \banyanE\footnote{\banyanE\ is documented as \url{http://www.exoplanetes.umontreal.ca/banyan/banyansigma.php}.} \citep{2018ApJ...856...23G}, which utilizes astrometric and kinematic measurements to establish the membership probability of a star in 30 known NYAs, represented by 3D ellipsoidal models in $XYZ$ and $UVW$ space \citep{2018MNRAS.475.2955L, 2018ApJ...856...23G, 2019MNRAS.486.3434L}. In total, \inputsample\ stars, located at an average distance of $28.22 \pm 0.02$\,pc and a distance range of $9.719-83.99$\,pc, were identified as part of this sample, presented in Table~\ref{tab:sample} and hereafter referred to as the input sample. Over half of these stars are low-mass M dwarfs, and most show signs of youth such as high X-ray, UV, or H$\alpha$ emission, fast rotation, or lithium emission lines. Stars within the input sample were further vetted in order to flag and exclude possible unresolved multiple systems (see Section~\ref{sec:multiple_systems}) and kinematic outliers (see Section~\ref{sec:kinematic_outliers}).

\subsection{Stellar kinematics}
\label{sec:kinematics}

We sourced most parallax, proper motion, and absolute radial velocity data from the \gaiadr\ data products \citep{2022arXiv220800211G}, while \hip\ data were used for very bright stars \citep{1997AA...323L..49P}, for which astrometry is sometimes more precise in \hip\ than \gaiadr. We also used other kinematic measurements compiled from the literature, and we used an error-weighted average to combine all available measurements. This allows us to obtain more precise and reliable measurements in order to better confirm the membership of members of \bpmg\ with \banyanE\ and compute more precise backward Galactic orbits for every member, which will in turn yield a more precise kinematic age estimate and help limit contamination by multiple systems and kinematic outliers. In addition, reliable and precise astrometric and kinematic measurements are also essential to minimize the bias on traceback ages due to measurement errors (see Section~\ref{sec:sensitivity_kinematic_errors}). The average $XYZ$ Galactic positions and $UVW$ space velocities of members of the input sample are presented in Table~\ref{tab:parameters}. The astrometric and kinematic measurements of all members of the input sample of \bpmg\ are presented in Table~\ref{tab:observations}.

\tablewidth{\textwidth}
\tabletypesize{\small}
\begin{deluxetable}{lc}[t]
\tablecolumns{2}
\tablecaption{Parameters of the input and core samples of \bpmg}
\label{tab:parameters}
\tablehead{\colhead{Parameter} & \colhead{Value}}
\startdata
\sidehead{\textbf{Input Sample}}
Number of stars & 76\\
Average distance & $28.22 \pm 0.02$\,pc\\
Distance range & $9.719-83.99$\,pc\\
$XYZ$ & ($22.14$, $-5.20$, $-16.71$)\,pc\\
$\sigma_{XYZ}$ & ($32.97$, $14.08$, $8.61$)\,pc\\
$UVW$ & ($-10$, $-15$, $-9$)\,\kms\\
$\sigma_{UVW}$ & ($8$, $5$, $5$)\,\kms\\
\sidehead{\textbf{Core Sample}}
Number of stars & 25\\
Average distance & $29.587 \pm 0.006$\,pc\\
Distance range & $9.719-71.55$\,pc\\
$XYZ$ & ($22.691$, $-4.308$, $-18.492$)\,pc\\
$\sigma_{XYZ}$ & ($29.698$, $13.940$, $8.106$)\,pc\\
$UVW$ & ($-10.2$, $-15.7$, $-8.64$)\,\kms\\
$\sigma_{UVW}$ & ($1.5$, $0.6$, $0.76$)\,\kms\\
\enddata
\tablecomments{The average $XYZ$ Galactic position and dispersion ($\sigma_{XYZ}$) and the average $UVW$ space velocity and dispersion ($\sigma_{UVW}$) are given for members of the input and core samples of \bpmg.}
\end{deluxetable}


\subsection{Unresolved multiple systems}
\label{sec:multiple_systems}

A large fraction of stars in the Galaxy are in fact part of multiple systems, for which the added stellar motion around the barycenter of the system, along the line of sight, modulates radial velocity measurements. Unresolved multiple systems can therefore appear as false positives when one attempts to select NYA members based on kinematic data if a stars' $UVW$ space velocity happens to match the kinematic model of an NYA.

Fortunately, it is possible to flag these false positives by comparing radial velocity measurements at different epochs. Indeed, the radial velocity of stars in multiple systems changes periodically, and inconsistent measurements can thus be used as a way to flag unresolved multiple systems. Therefore, in order to exclude kinematic measurements contamination by the presence of these systems, we selected members from the input sample of \bpmg\ for which at least two radial velocity measurements are available in the literature and did not display variations $\geq 0.6$\,\kms, a value larger than the typical radial velocity scatter in our data set, indicating the possible presence of an unresolved companion. Also excluded were known spectral binaries from the literature and the Washington Double Star (WDS) catalog, stars with excess brightness in the color-magnitude diagram (CMD), as well as stars with a \gaiadr\ Renormalized Unit Weight Error (RUWE) $\geq 2.0$.\footnote{The RUWE is documented in the Gaia Data Release Documentation at \url{https://gea.esac.esa.int/archive/documentation/GDR2/Gaia_archive/chap_datamodel/sec_dm_main_tables/ssec_dm_ruwe.html}.} This statistical indicator is a way to assess the reliability of astrometric data. It is expected to be $\sim 1.0$ for sources for which the single-star astrometric solution is a good fit \citep{2018LL-LL-124L} while a RUWE $> 1.4$ is a sign of an unresolved binary system \citep{2021ApJ...907L..33S, 2021MNRAS.506.2269E}. We chose a slightly less robust threshold because we also require stars to show relatively stable radial velocities over time. The various multiplicity indicators of members of the input sample of \bpmg\ are presented in Table~\ref{tab:multiple}.

The result of this selection process is a clean, uncontaminated subset of the input sample of \bpmg\ made up of \coresample\ stars located at an average distance of $29.587 \pm 0.006$\,pc and a distance range of $9.719-71.55$\,pc, hereafter referred to as the core sample of \bpmg, which can be used to compute the traceback age of the association. The average $XYZ$ Galactic positions and $UVW$ space velocities of members of the core sample are presented in Table~\ref{tab:parameters}. We note that all parameters of the input and core samples are similar, apart from the number of stars included in the sample and the $UVW$ space velocity dispersion ($\sigma_{UVW}$), which is significantly tighter for the core sample, as expected. The remaining \extendedsample\ stars excluded from the core sample by this selection process, hereafter referred to as part of the extended sample, are not considered in the computation of the traceback age. The following stars deserve additional discussion:

\begin{description}
\item[HD 165189] This star is a slight kinematic outlier with only three available radial velocity measurements. It is more than 1 magnitude brighter than other members of \bpmg\ in the CMD, and the same is true of its companion located at a 1" separation. The absolute V vs. V--J is comparable to \cite{2015MNRAS.454..593B} and seems to best match the \gaiadr\ parallax. Therefore, the \gaiadrtwo\ and EDR3 photometry might be unreliable. For now, it remains a  \bpmg\ candidate and was kept in the extended sample.
\item[HD 207043] This star's position in a \gaiadr\ CMD is inconsistent with other members of \bpmg, which puts into question its membership in the association. For now, it remains a \bpmg\ candidate and was kept in the extended sample.
\item[AF Psc] With updated kinematics, this triple star system with separations of 19" and 1000" \citep{2014ApJ...792...37M, 2014AJ....147..146K, 2020AA...642A.179M} is a poor kinematic match to the \banyanE\ model of \bpmg. As a result, it was excluded from the extended sample.
\item[G 271--110] With updated kinematics, this star is a poor kinematic match to the \banyanE\ model of \bpmg, and its position in a CMD is also fainter than other members of \bpmg. \cite{2014ApJ...792...37M} noted that it also does not show a lithium absorption, whereas it normally should at its temperature if it were a member of \bpmg. Also, at $1.71$, its \gaiadr\ RUWE is a bit high, so it may be an unresolved binary system. As a result, it was excluded from the extended sample.
\item[HD 14082 A] This star displays slight radial velocity variations. It was excluded from the extended sample, and its co-mover, HD 14082 B, located at a separation of 14", was used instead.
\item[HD 173167] This star is a spectroscopic binary system with few radial velocity measurements that display slight variations \citep{2016MNRAS.459.4499E}. This star was excluded from the extended sample, and its co-mover, Smethells 20, was used instead.
\end{description}

\begin{figure}[t]
    \includegraphics{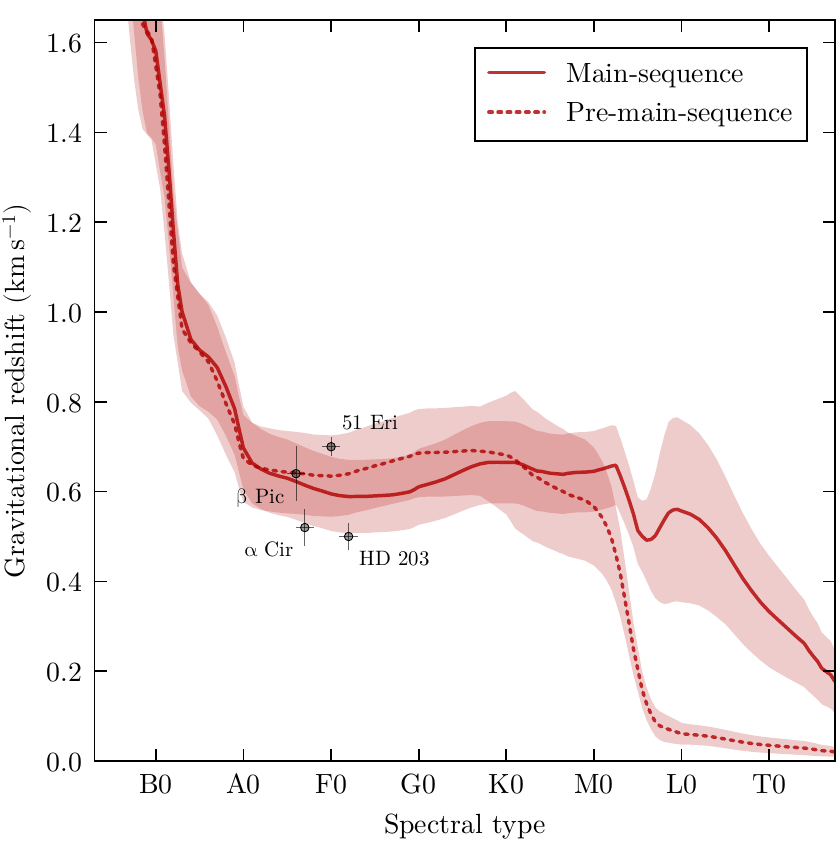}
 	\caption{Gravitational redshift as a function of spectral type based on our determinations of stellar masses and radii from empirical mass--spectral type and radius--spectral type sequences for main-sequence (solid red line) and pre-main-sequence objects (dotted red line). The shaded areas show the uncertainties based on individual spectral type measurement errors and propagated standard deviations of empirical mass and radius sequences. For most late-type members of \bpmg, the expected absolute radial velocity shifts due to the gravitational redshift alone are $\sim 0.6$\,\kms. For comparison, the radial velocity shifts due to the gravitational redshift of four stars within \bpmg\ that benefit from interferometric radius measurements are indicated by black circles and error bars \citep{2004IAUS..219...80K, 2008MNRAS.386.2039B, 2011ApJ...743..158S}.}
 	\label{fig:grav_redshift}
\end{figure}

\subsection{Gravitational redshift}
\label{sec:grav_redshift}

As a result of general relativity, radial velocity measurements are biased due to the gravitational redshift as light emitted at the surface of a star leaves its gravitational potential. Light is also later blueshifted slightly by the Sun's and the Earth's gravitational potentials, although this effect is negligible in comparison and safely ignored. The gravitational redshift is given as
\begin{equation}\label{eq:grav}
1 + z_{grav} = \frac{\lambda_{obs}}{\lambda} = {\left( 1 - \frac{2GM}{c^2 R} \right)}^{-\slfrac{1}{2}}.
\end{equation}

\noindent
A star's radial velocity (RV) is obtained indirectly by measuring its Doppler shift, as it moves away from or toward the observer. It is estimated by its non-relativistic expression:
\begin{equation}\label{eq:doppler}
1 + z_{Doppler} = \frac{\lambda_{obs}}{\lambda} = 1 + \frac{RV}{c}.
\end{equation}

\noindent
By combining equations~\ref{eq:grav}~and~\ref{eq:doppler}, the expected bias on traceback ages due to the gravitational redshift of a star can be estimated as
\begin{equation}\label{eq:rv_bias}
\Delta RV_{grav} = c \left( {\left( 1 - \frac{2GM}{c^2 R} \right)}^{-\slfrac{1}{2}} - 1 \right).
\end{equation}

\noindent
Therefore, a star will appear to have a greater radial velocity (i.e., it will appear to move away faster from the observer). From equation~\ref{eq:rv_bias}, this bias can be derived by two fundamental parameters: the stellar mass ($M$) and radius ($R$). These values, however, cannot be measured directly for most stars, and an approximation based on spectral type is used instead (see Figure~\ref{fig:grav_redshift}). 
This approximation is calibrated on interferometric radius measurements of main-sequence stars \citep{2016AA...586A..94L}, semi-empirical spectral energy distribution (SED)-based determinations of angular radii of members of the core sample of \bpmg\ by \cite{2013ApJS..208....9P}, updated with \gaiadr\ parallax measurements, as well as ensembles of dynamical masses of eclipsing binaries for young and pre-main-sequence stars compiled by \cite{2004ApJ...604..741H} and other sources, updated with \gaiadr\ parallax measurements when relevant. We avoid basing our approximation on absolute photometry or bolometric luminosity, due to the potentially large bias that we would obtain for yet unknown unresolved multiple systems in the core sample. 

While the impact is relatively small, \rvshift\ for most stars, traceback analysis is highly sensitive to such biases: because members of the closest NYAs are spread across a large fraction of the sky, this bias will create the illusion that members are moving away from the observer in multiple directions, causing them to appear to reach a minimal spatial extent at a more recent epoch when performing a traceback analysis. More distant NYAs, which do not cover a large fraction of the sky, are proportionally less affected by this bias.

This bias also affects the determination of membership probabilities for NYAs with \banyanE, because this process also utilizes radial velocities as one of its input parameters, although the impact is much less pronounced, given that the inherent velocity dispersion of young associations is larger, at $\sim 1-3$\,\kms.

\begin{figure*}[!ht]
    \includegraphics{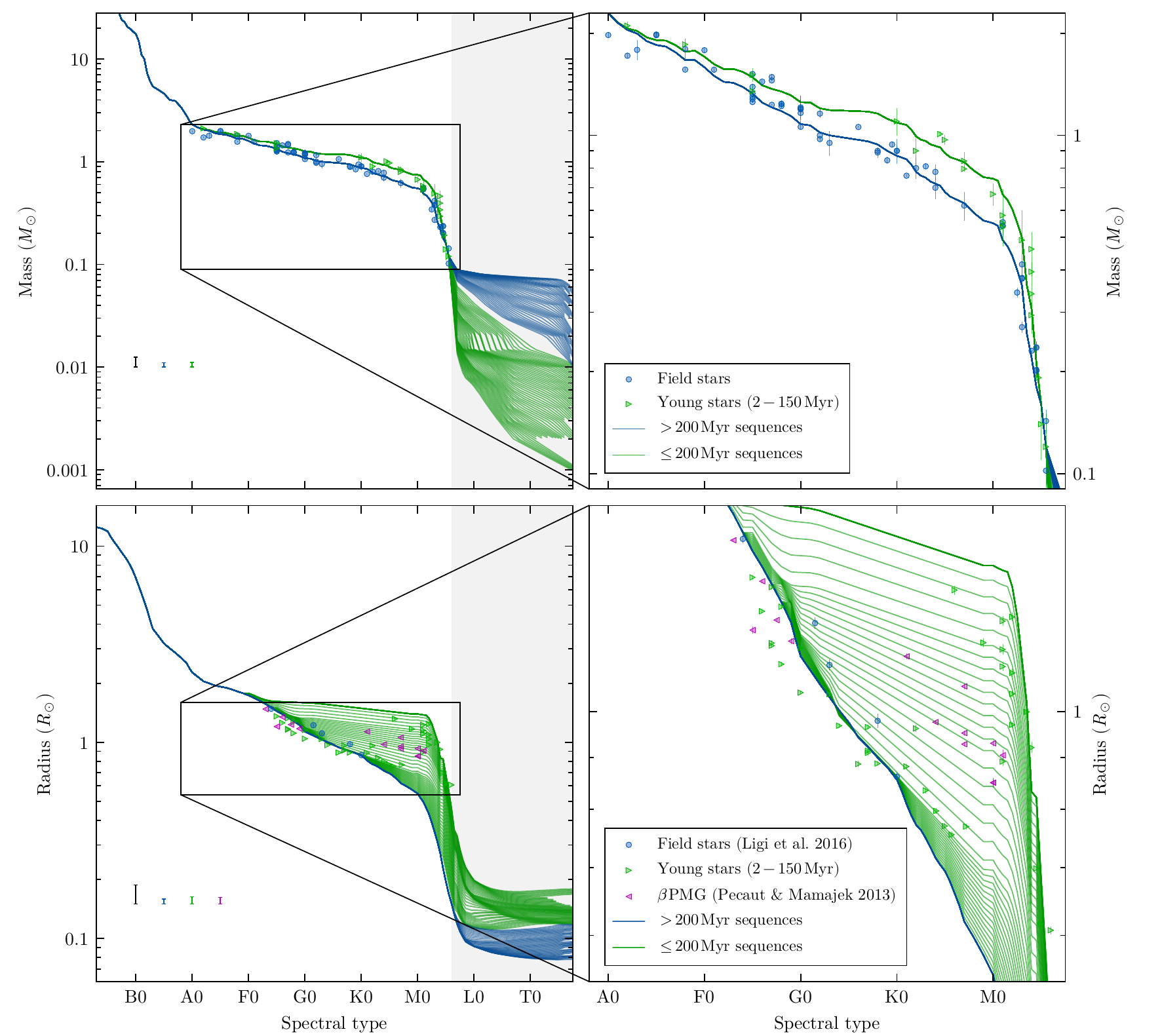}
 	\caption{Stellar masses (top panels) and radii (bottom panels) as a function of spectral type for field (blue circles) and young stars ($2-150$\,Myr, green triangles). The masses of both field and young stars were dynamically measured \citep{2004ApJ...604..741H, 2015ApJ...813L..11M, 2016AJ....152..175N, 2017AA...607A..10A, 2018AA...618A..23R,  2018AA...620A..33J, 2019ApJ...884...42S, 2021ApJ...908...46B, 2021ApJ...908...42P}. The radii of field stars were measured by interferometry \citep{2016AA...586A..94L} while the radii of young stars were derived from their SED using the modified P13 spectral type--radius relation \citep{2013ApJS..208....9P}, which reproduces empirical measurements at young ages, representative of \bpmg\ (magenta triangles). Blue ($> 200$\,Myr) and green ($\leq 200$\,Myr) lines represent our empirical spectral type--mass and spectral type--radius sequences with model extrapolations for spectral types later than M5 (light gray shaded area), for ages ranging from $1$\,Myr to $10$\,Gyr, using $50$ steps of $0.08$\,dex in log space. The discontinuity at $200$\,Myr is due to the adoption of distinct spectral type to effective temperature sequences as prescribed by \cite{2013ApJS..208....9P}. The error bars represent the typical uncertainty associated with our sequences for spectral types M5 and earlier (same colors), and the black error bar represents the uncertainty assigned to model extrapolations.}
 	\label{fig:masses_radii}
\end{figure*}

\subsubsection{Stellar masses}
\label{sec:stellar_masses}

The mass of members of the core sample of \bpmg\ was estimated using their spectral type in order to avoid biases in cases of potential unresolved multiple systems that may remain in this sample. We used the \cite{2013ApJS..208....9P} spectral type--mass relation\footnote{See \url{https://www.pas.rochester.edu/~emamajek/EEM_dwarf_UBVIJHK_colors_Teff.txt}.} (P13 relation hereafter) and compared it to dynamical mass measurements of known A2--M5 young stars ($2-150$\,Myr) from the literature, improved with \gaiadr\ parallax measurements when relevant. Figure~\ref{fig:masses_radii} shows that the dynamical masses of young stars are slightly larger than those of old field stars at a fixed spectral type. This is expected, given that stars warm up slightly in their pre-main-sequence phase, which lasts for $20-100$\,Myr for $\leq 1$\,\msol\ stars (e.g., see \citealp{2016ApJ...823..102C}).

We built a young version of the P13 spectral type--mass relation by adjusting a two-segment linear fit to the difference in masses between the P13 relation and measured dynamical masses in log space to shift the P13 relation upward slightly, as shown in Figure~\ref{fig:masses_radii}. We found a relative standard deviation of $11\,\%$ for the dynamical masses of young stars with respect to our modified young sequence, which we adopt as $1 \sigma$ uncertainties on our spectral-type-based mass estimations.

\subsubsection{Stellar radii}
\label{sec:stellar_radii}

We similarly used the P13 spectral type--radius relations to determine the radii of the stars in the core sample based on their spectral types, to avoid potential biases due to unresolved binaries. We modified the main-sequence version of the P13 relations using the sample of known members of \bpmg\ in P13 to obtain a spectral type--radius relation that is appropriate for the younger age of \bpmg, at a phase when stars are known to have an inflated radius due to their slow contraction during the pre-main-sequence phase (e.g., see \citealp{1997ApJ...491..856B}), which is slowed down significantly in the case of low-mass stars by strong magnetic fields driven by their fast rotations (e.g., see \citealp{2014ApJ...792...37M}). P13 measured angular semi-diameters based on an SED fitting method \citep{2006AA...450..735M}, which we translated into stellar radii using the best available parallax measurements, mostly based on \gaiadr. We calculated a relative standard deviation of $9\,\%$ for the radius measurements with respect to our modified spectral type--radius sequence, which we adopt as our measurement error. We show the resulting sequence in Figure~\ref{fig:masses_radii}.

For both the masses and radii, we used the \cite{1997ApJ...491..856B} evolutionary models to extrapolate our empirical sequences beyond spectral type M5, and we assigned larger relative measurement errors of $20\,\%$ to account for possible model systematics, which may be especially important for young low-mass stars.

By combining the spectral type--mass and spectral type--radius relations described above, we built a spectral type--gravitational redshift relation shown in Figure~\ref{fig:grav_redshift}. The application of this correction to radial velocity measurements of members of the core sample of \bpmg\ is presented in Table~\ref{tab:rv_shift}.

\begin{figure}
    \includegraphics{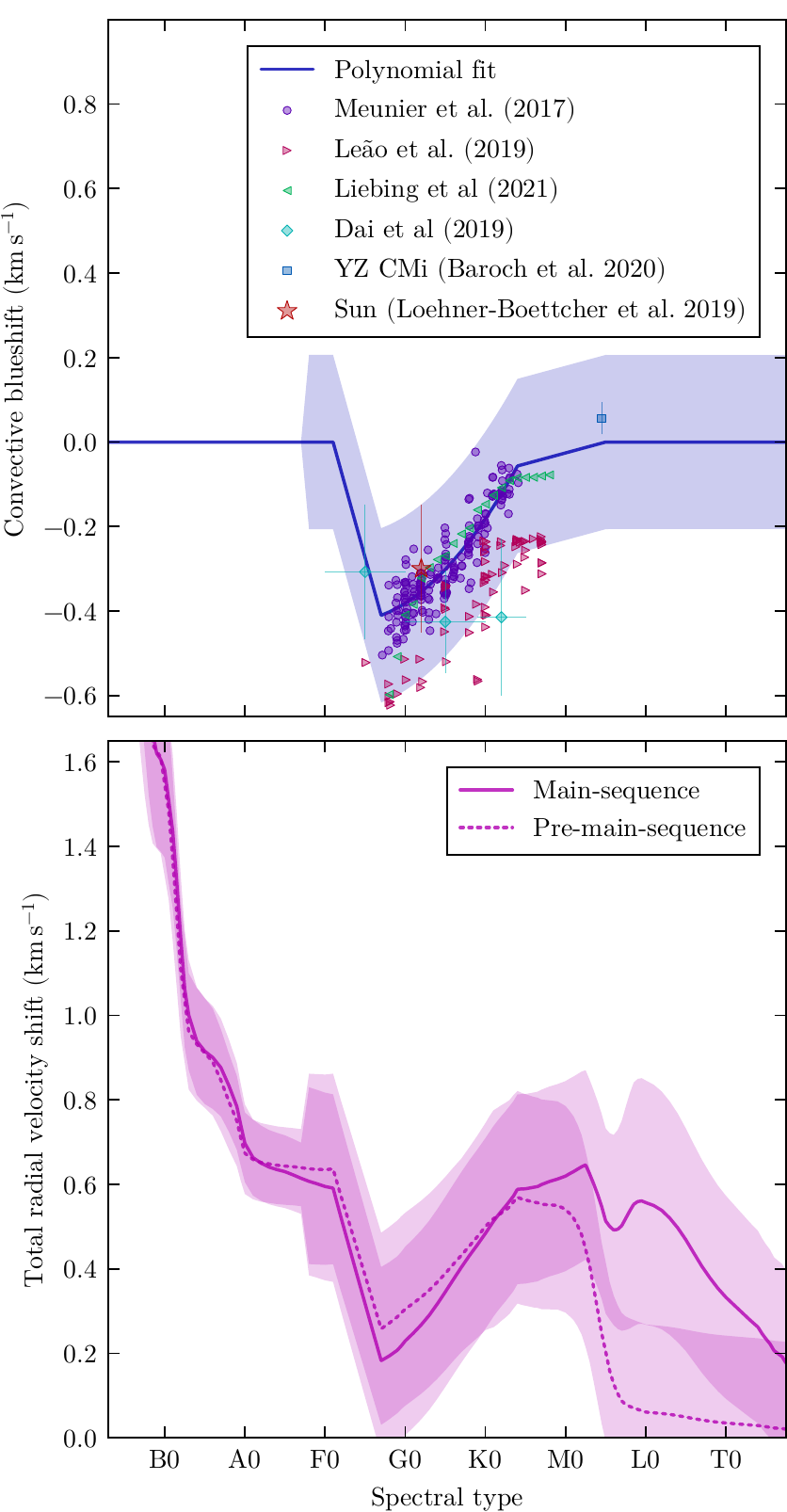}
 	\caption{Convective blueshift (top panel) and total absolute radial velocity shift (bottom panel), including the effects of the gravitational redshift for main-sequence (solid purple line) and pre-main-sequence objects (dotted purple lines), as a function of spectral type. The shaded areas show the uncertainties based on individual spectral type measurement errors and propagated standard deviations of empirical mass and radius sequences. The convective blueshift three-segment polynomial sequence (solid blue line) was fitted with data from \cite{2017AA...607A.124M}, and the uncertainty was chosen to be wide enough to encompass data from: \cite{2019MNRAS.483.5026L}, \cite{2021AA...654A.168L}, \cite{2019ApJ...871..119D}, \cite{2020AA...641A..69B}, \cite{2019yCat..36240057L}, \cite{2013AA...550A.103A}, and \cite{1988AJ.....96..198G}, including measurements for YZ~CMi, a member of \bpmg, and the Sun.}
    \label{fig:conv_blueshift}
\end{figure}

\subsection{Convective blueshift}
\label{sec:conv_blueshift}

With the presence of convection cells at the stellar surface, light is emitted from areas moving toward or away from the observer, resulting in a Doppler broadening of spectral lines. Because rising gas is hotter and brighter, and it accounts for a larger fraction of the stellar surface, the broadening is uneven and causes a change in the shape of spectral lines and a net blueshift of the spectrum.

Convective blueshift has historically been difficult to measure. By comparing precise astrometric radial velocity measurements from the \hip\ and \gaia\ missions with spectroscopic measurements, it is possible to isolate phenomena, other than the Doppler effect, that can shift spectral lines \citep{2019MNRAS.483.5026L}. Convective blueshift can also be measured by studying the shape of the differential of spectral lines \citep{2017AA...607A.124M}. Its impact on the instrumental point spread function is dependent on the instrumental resolution, the measurement method, and the spectral range. Lower stellar mass and higher stellar activity are linked to lower levels of convective blueshift, while lower metallicity and higher effective temperature are linked to higher levels. On the other hand, there is little effect from age or cyclic activity \citep{2021AA...654A.168L}. The net radial velocity shifts derived by cross-correlation also depend on the spectral resolution of the instrument (i.e., the lower the resolution, the smaller the sensitivity to changes in the shapes of spectral lines) and the wavelength range \citep{2013AA...550A.103A}.

The effect of convective blueshift is smaller than the aforementioned gravitational redshift, so the sum of both effects is expected to be a redshift (see Figure~\ref{fig:conv_blueshift}). For spectral types earlier than F2, we assume no convective blueshift, due to the absence of convection cells at the stellar surface. A three-segment polynomial sequence was fitted to blueshift measurements from the literature for F2--F7, F7--K4, and K4--M5 stars with a conservative estimated error of $0.2$\,\kms, to account for the dependence of the impact of convective blueshift on the instrumental point spread function. For stars later than M5, although they are expected to be fully convective, no robust measurements of convective blueshift are available yet. The case of YZ~CMi \citep{2020AA...641A..69B}, a member of \bpmg, is consistent with a null convective blueshift within the error bars. For this reason, we adopt a null correction with a $0.2$\,\kms\ measurement error until more data are available. The application of this correction to radial velocity measurements of members of the core sample of \bpmg\ is presented in Table~\ref{tab:rv_shift}.

\section{Traceback analysis}
\label{sec:traceback_method}

Traceback analysis is performed with the custom Python package \texttt{kanya}\footnote{\texttt{kanya}: Kinematic Age for Nearby Young Associations. The source code is available at \url{https://github.com/***/kanya}.}, which is capable of computing the traceback age of a given sample of stars. The backward Galactic orbit of each star is computed and the spatial extent of the association is determined using a range of association size metrics (see Section~\ref{sec:metrics}) at every temporal step, in order to identify the epoch of minimal association size. As will be demonstrated in this section, this method is sensitive to sample contamination, biases introduced by measurement errors, and systematic effects that impact absolute radial velocity measurements. It is therefore crucial to address each of these potential issues in order to determine reliable traceback ages.

\subsection{Backward orbital integration}
\label{sec:orbit_intergration}

A correction to radial velocity measurements was first applied before computing backward Galactic orbits, in order to compensate for the effects of gravitational redshift and convective blueshift (see sections~\ref{sec:grav_redshift}~and~\ref{sec:conv_blueshift}). Then the full current-day 6D kinematics of members of the core sample of \bpmg\ are computed (see Table~\ref{tab:kinematics}). We use the galpy\footnote{galpy is documented at \url{http://github.com/jobovy/galpy}.} Python package \citep{2015ApJS..216...29B} with the Galactic potential model I from \cite{2013AA...549A.137I} to compute independent backward Galactic orbits for every member of the NYA. This model, which is an updated version of the Galactic potential from \cite{1991RMxAA..22..255A}, assumes a $9.5 \times 10^9$\,\msol\ spherical bulge, a $6.6 \times 10^{10}$\,\msol\ \cite{1975PASJ...27..533M} disk, and a $1.8 \times 10^{12}$\,\msol\ spherical halo. This is the same Galactic potential used by \cite{2020AA...642A.179M}, who concluded that variations in the kinematic age caused by the choice of the Galactic potential are smaller than the main source of uncertainty, due the short integration time for such young NYAs.

$XYZ$ Galactic coordinates are transformed from a heliocentric, right-handed Cartesian system into a Galactocentric, left-handed cylindrical system \citep{2020MNRAS.499.5623Q}. The following values for the Sun's current-day Galactic position are adopted for this transformation:
\begin{align}\label{eq:sun_position}
R_\odot &= 8.12 \pm 0.03\textrm{\,kpc}\notag\\
\phi_\odot &= 0\textrm{\textdegree}\\
z_\odot &= 5.6 \pm 5.8\textrm{\,pc},\notag
\end{align}

\noindent
where \rsol\ is the Sun's Galactocentric radius \citep{2018AA...615L..15G} and $z_\odot$ is the Sun's height above or below the Galactic plane \citep{2019ApJ...885..131R}. The Sun's Galactocentric longitude $\phi_\odot$ is null by definition. We adopt the following peculiar Solar motion to transform $UVW$ space velocities \citep{2010MNRAS.403.1829S}:
\begin{align}\label{eq:sun_peculiar_velocity}
U_\odot &= 11.1_{-0.8}^{+0.7}\textrm{\,\kms}\notag\\
V_\odot &= 12.2 \pm 0.5\textrm{\,\kms}\\
W_\odot &= 7.3 \pm 0.4\textrm{\,\kms}.\notag
\end{align}

\noindent
We also adopt the following local standard of rest (LSR) rotational velocity \citep{2010MNRAS.403.1829S}:
\begin{equation}
V_{\textrm{LSR}} = 233 \pm 1.4 \textrm{\,\kms}.
\end{equation}

\noindent
Once Galactic orbits are computed for every member of the NYA, their individual Galactocentric positions and velocities are transformed back into $XYZ$ Galactic coordinates and $UVW$ space velocities.

Though this approach is more physically accurate than simply ignoring the Galactic potential's impact on past trajectories, its inclusion may have little impact on both the bias and error of our method, for an association as young as $\sim 24$\,Myr (see Figure~\ref{fig:simulated_samples}). Members of NYAs have similar $XYZ$ Galactic positions and $UVW$ space velocities, and thus they follow similar Galactic orbits as well. Because only the members' relative positions to the average position of the NYA are considered, the Galactic potential may have little impact on traceback ages.

\subsection{Kinematic outliers}
\label{sec:kinematic_outliers}

Stars in the core sample (see Section~\ref{sec:multiple_systems}) were further investigated in order to identify kinematic outliers. First, the $XYZ$ Galactic position and $UVW$ space velocity standard deviations are computed independently along each 6D component at every temporal step of the stars' trajectory. Stars that stray away from the core of the NYA, beyond a $3\sigma$ threshold in either $XYZ$ Galactic position or $UVW$ space velocity at any point and in any component, are flagged as kinematic outliers and excluded from the computation of size metrics. This process is repeated recursively until no more stars are flagged as kinematic outliers. 

The robust covariance estimator developed by \cite{2011JMLR...12.2825P} as part of the scikit-learn\footnote{Scikit-learn's robust covariance estimator is documented at \url{https://scikit-learn.org/stable/modules/covariance.html?highlight=robust\%20covariance\#robust-covariance-estimation}.} Python package was then used to identify kinematic outliers independently at every temporal step of the traceback. Stars identified as outliers at least $70\,\%$ of the time along their trajectory were flagged as kinematic outliers and ignored in the computation of all association size metrics. Scikit-learn's robust covariance estimator uses the empirical covariance matrix ($\Sigma$) of the stars' $XYZ$ Galactic positions and $UVW$ space velocities:

\begin{equation}
\label{eq:covariance_matrix}
\Sigma = E \left( (\boldsymbol{x} - E(\boldsymbol{x})) (\boldsymbol{x} - E(\boldsymbol{x}))^T \right),
\end{equation}

\noindent
where $E(\boldsymbol{x})$ is the expected value of a random vector $\boldsymbol{x}$. Since the empirical covariance matrix is sensitive to the presence of outliers in the data, it is necessary to perform outlier detection in order to compute a robust estimator of the covariance matrix. This is accomplished by computing the Mahalanobis distance \citep{1936PNISI...2M} of each star. For an observation $x_i$ of a distribution with an empirical covariance matrix $\Sigma$ and a mean $\mu$, the Mahalanobis distance ($d_M$) to the mode is given by

\begin{equation}
\label{eq:Mahalanobis}
d_M^2(x_i) = (x_i - \mu)^T \Sigma^{-1} (x_i - \mu).
\end{equation}

\noindent
This approach is an effective tool to perform outlier detection in noisy and irregular data sets because it takes into account covariances in the distribution. In summary, the distance is measured and scaled to the standard deviation along each principal component of the 6D distribution.

The number of stars included in the sample is a limitation to the precision and accuracy of the traceback method. As one would expect, traceback ages are more precise and accurate with a larger sample. However, as more stars are included in the sample, some will likely be kinematic outliers or unrecognized binary stars, and these will in turn both negatively affect the precision of traceback ages and create an additional bias toward younger ages. The trajectories of these stars tend to deviate significantly from the average trajectory of the association, and even just a few such outliers can dramatically lower the traceback age by artificially increasing the dispersion in $XYZ$ Galactic positions and $UVW$ space velocities. Therefore, it is advantageous to limit traceback samples to smaller ensembles of stars, free of such contamination. However, including more stars in the sample will decrease the minimal theoretical error of ages determined with the traceback method, because the epoch of minimal association size will be more clearly defined, provided that no kinematic outliers affect the results negatively.

\subsection{Association size metrics}
\label{sec:metrics}

The empirical covariance matrices of members of an NYA were calculated at each temporal step in both $XYZ$ or \xez\ Galactic coordinate systems. The latter, identical to the system used by \cite{2020AA...642A.179M}, is a heliocentric curvilinear coordinate system that minimizes variations along every individual component during the backward orbital integration by moving the origin on a circular orbit around the Galactic Center at a frequency of $\omega_\odot = V_{\textrm{LSR}} / R_\odot = 28.7 \textrm{\,km} \textrm{\,s}^{-1} \textrm{\,kpc}^{-1}$. In addition to the individual diagonal terms of the empirical covariance matrices, the determinant and the trace of the associated covariance matrices were investigated as plausible association size metrics, similarly to the approach taken by \cite{2020AA...642A.179M}.

We also considered the spatial--kinematic cross-covariance terms for members at each temporal step as potential association size metrics, in both $XYZ$ and \xez\ Galactic coordinate systems. Covariances between spatial and kinematic terms of a given direction are expected to be minimal at the epoch of stellar formation and naturally increase over time (e.g., see \citealp{2019MNRAS.489.3625C}). Once again, the determinant, trace, and individual diagonal terms of the cross-covariance matrices were included in the size metrics under consideration.

We implemented both “empirical” and “robust” versions of each association size metric based on the empirical $XYZ$ and \xez\ covariance matrices (equation~\ref{eq:covariance_matrix}). While empirical metrics give each member an equal weight, robust metrics assign them a weight inversely proportional to their Mahalanobis distance (see equation~\ref{eq:Mahalanobis}) to the association's core.

\begin{figure*}
 	\includegraphics{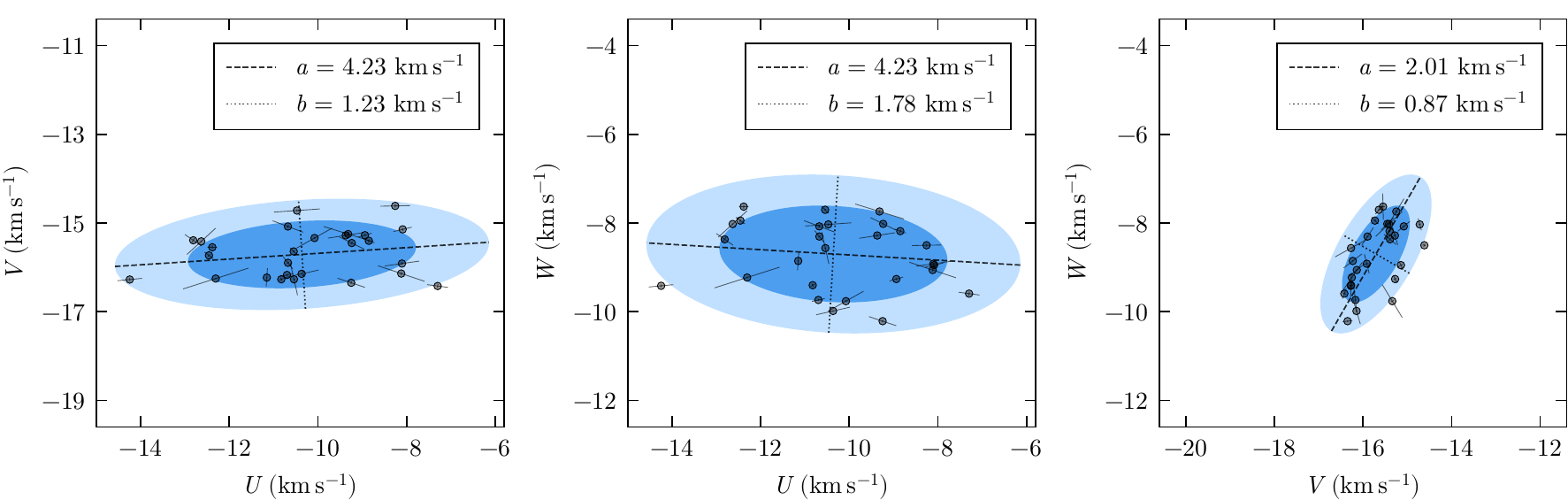}
 	\caption{Application of the extreme deconvolution algorithm developed by \cite{2011AnApS...5.1657B} on the current-day $UVW$ space velocities of members of the core sample of \bpmg\ in order to mitigate the effect of the elongation of error bars along the line of sight due to relatively larger measurement errors on absolute radial velocity measurements. The dark and light blue shaded areas respectively represent the 67\,\% and 95\,\% probability ellipsoids defined by the diagonal terms of the covariance matrix. The length of the semi-major (a) and semi-minor (b) axes of the 95\,\% probability ellipsoids are indicated in the legends.}
 	\label{fig:error_ellipse_uvw}
\end{figure*}

We also investigated the use of median absolute deviations (MADs) as an alternate set of association size metrics, given that the MAD depends more weakly on a small number of outliers. We computed the MAD along each spatial component, in both $XYZ$ and \xez\ Galactic coordinate systems, as a way to represent the spatial extent of an NYA at each epoch. This metric is not only less sensitive to kinematic outliers than the standard deviation, it is also expected to be even less sensitive to outliers than “robust” metrics described above, because the latter still assign a small but nonzero weight to significant outliers.

Finally, we computed the minimum spanning tree (MST) of the position of members at each epoch in both $XYZ$ and \xez\ Galactic coordinate systems using a Kruskal algorithm \citep{1956PAMS....7...1K}. MSTs are undirected graphs that link all vertices with the minimal possible total edge length, without allowing for loops. The mean edge length and the MAD of the edge lengths of the MST are used as association size metrics, which remain valid even in highly non-Gaussian geometries. This approach has the advantage of determining a characteristic size while being less sensitive to the shape of the NYA. A robust version of the mean edge length of the MST, using the same weights as other robust size metrics, was also explored.

\subsection{Error on the traceback age}
\label{sec:error_traceback_age}

The traceback age is highly dependent on which stars are included in the computation of association size metrics. To assess this error on the traceback age due to the sensitivity to sample definition, we employed a jackknife Monte Carlo approach. We used $1000$ iterations, each made up of a $50\,\%$ fraction of randomly selected stars the sample, to measure the standard deviation on the traceback age for every size metric.

We also made use of a Monte Carlo approach to compute the error on the traceback age due to the measurement errors in the astrometric and kinematic data. Randomized Gaussian fluctuations with a standard deviation equal to the reported measurement errors were added to radial velocity, proper motion, and parallax measurements for $1000$ iterations. The total error on the traceback age was computed with a Monte Carlo approach including both the jackknife and measurement errors.

\subsection{Simulated samples}
\label{sec:simulated_samples}

We constructed simulated NYA samples, designed to best represent the core sample of current-day members \bpmg\ in order to test and characterize our method precision, accuracy, and sensitivity to several sources of bias.
A model star was initialized with an $XYZ$ Galactic position and a $UVW$ space velocity equal to the current-day average values of members of the core sample of \bpmg\ (see Table~\ref{tab:parameters}). This model star's trajectory was traced back $24$\,Myr to the birth epoch, a length of time similar to recent age estimates of \bpmg\ using the isochrones and LDB methods \citep{2014ApJ...792...37M}. Then, a randomized Gaussian sample of \coresample\ synthetic stars, the same number as the number of stars in the core sample of \bpmg, was created, with average $XYZ$ Galactic positions and $UVW$ space velocities equal to those of the model star at the birth epoch, an initial $XYZ$ Galactic position scatter of $3.0$\,pc, and an initial $UVW$ space velocity scatter of $1.0$\,\kms, along all three axes. It is impossible to know for sure what the initial $XYZ$ Galactic position scatter of \bpmg\ was. However, the value used is similar to the current-day scatter along the $\zeta^\prime$-axis, an axis with very little growth over time. The initial $UVW$ space velocity scatter was approximated by the average current-day $UVW$ space velocity scatter of members of the core sample of \bpmg, because we assume that $UVW$ space velocities are relatively unchanged since the epoch of stellar formation for such a young NYA.

We used the extreme deconvolution algorithm\footnote{The extreme deconvolution algorithm is documented at \url{https://github.com/jobovy/extreme-deconvolution}.} developed by \cite{2011AnApS...5.1657B} to mitigate the impact of elongation of $UVW$ space velocities along the line of sight, due to larger errors on radial velocity measurements than proper motion measurements. This general algorithm can infer a $d$-dimensional distribution function from a set of heterogeneous and noisy samples, and it can treat uncertainties properly. When applied on the current-day $UVW$ space velocities of members of the core sample of \bpmg\ and their respective uncertainties (see Figure~\ref{fig:error_ellipse_uvw}), the resulting ellipsoids offer a more accurate description of distribution of $UVW$ space velocities in the core sample of \bpmg, which in turn is used to create more representative simulated samples.

\begin{figure*}[!ht]
 	\includegraphics{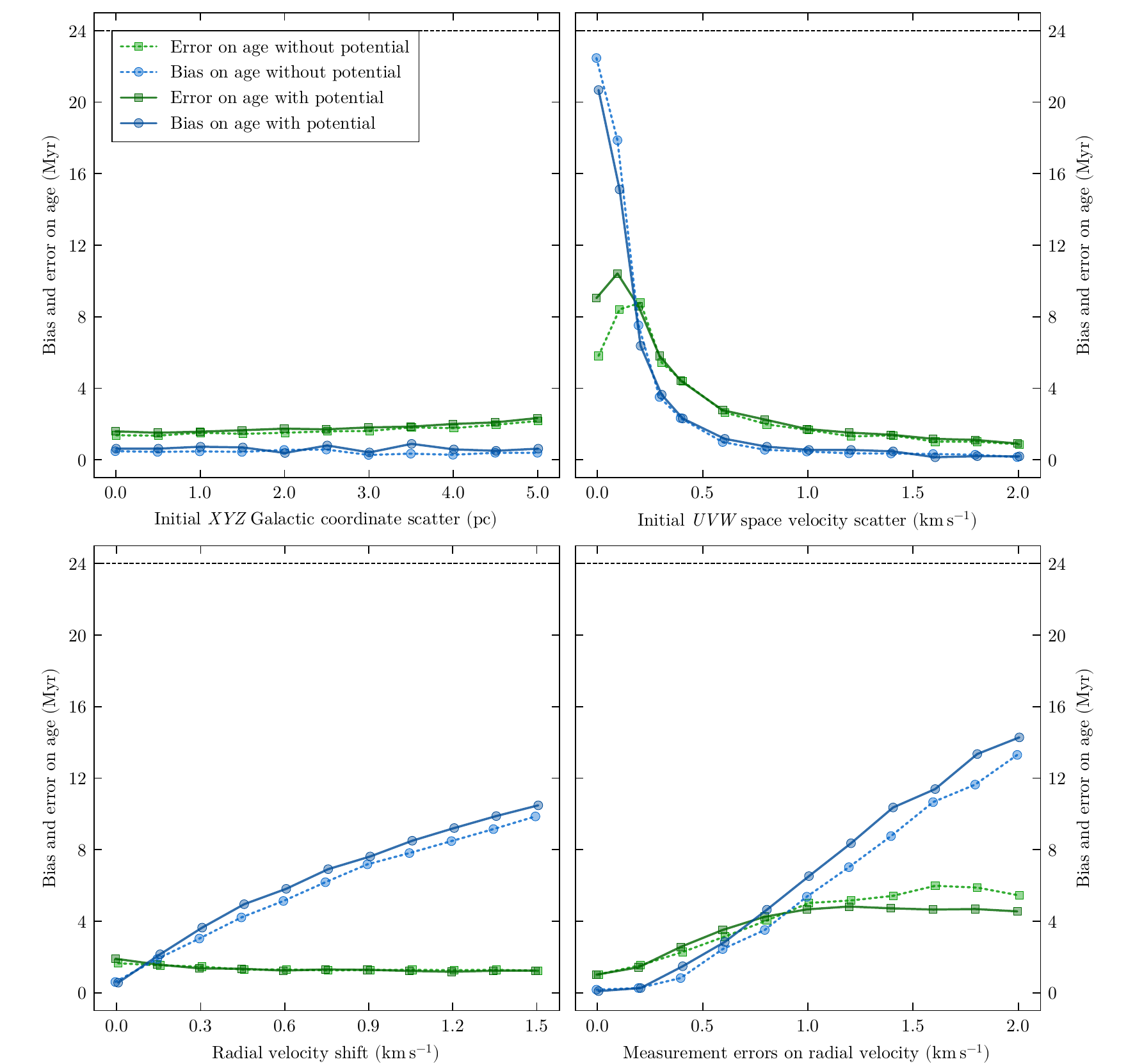}
 	\caption{Bias (blue curves and circles) and error (green curves and squares) on the traceback age as a function of the initial $XYZ$ Galactic position scatter, the initial $UVW$ space velocity scatter, the radial velocity shift and measurement errors on radial velocity measurements, for $24$\,Myr old simulated NYAs made up of $25$ members each, using the $\xi^\prime$ variance as the association size metric. The bias on age is defined as the average measured age offset of simulated NYAs with respect to their actual age ($24$\,Myr), and the error on age is the standard deviation of the measured ages of simulated NYAs. Dotted and solid curves respectively represent the bias and error on the traceback age with and without the Galactic potential model I from \cite{2013AA...549A.137I}. The gray dashed line shows the actual simulated NYA's age. Simulation parameters are set to match members of the core sample of \bpmg: measurement errors on radial velocity, proper motion, and parallax are equal to average measurement errors of members of the core sample of \bpmg, the initial $XYZ$ Galactic position scatter is set to $3.0$\,pc, the initial $UVW$ space velocity scatter is set to $1.0$\,\kms, and the bias on radial velocity is set to $0.0$\,\kms. The results of simulations without the Galactic potential are similar to the results with the Galactic potential taken into account or only slightly offset, which seems to confirm our assumption that taking into account the effect of the Galactic potential has little impact on the final traceback age.}
 	\label{fig:simulated_samples}
\end{figure*}

The simulated sample were then projected $24$\,Myr forward in time such that synthetic stars are located as near to the Sun as current-day members of \bpmg. $XYZ$ Galactic positions and $UVW$ space velocities were transformed into observables, and Gaussian-like fluctuations were added to the true sky coordinates, parallax, radial velocities, proper motions of each synthetic star in order to simulate measurement errors. A radial velocity shift was also added to simulate the bias on traceback ages due to the gravitational redshift and convective blueshift (see sections~\ref{sec:grav_redshift}~and~\ref{sec:conv_blueshift}). The errors on astrometric and kinematic measurements, and the radial velocity shifts were drawn from the real set of individual members of the core sample of \bpmg. The $24$\,Myr old simulated, biased, and noisy sample is then traced back over $50$\,Myr, and its size is computed with the full array of association size metrics (see Section~\ref{sec:metrics}) in order to find the epoch of minimal spatial extent.

\subsubsection{Sensitivity to kinematic measurement errors}
\label{sec:sensitivity_kinematic_errors}

An important bias to account for when computing the age of a coeval group of stars with traceback methods is the impact of Gaussian-like measurement errors in astrometric and kinematic data that may not simply result in a Gaussian-like distribution in the estimated epoch of minimal association size, but rather in a more complex probability distribution that will be not only inflated by an additional error term but also systematically biased with respect to the true epoch of minimal size closer to the current-day epoch.

The total observed Galactic position scatter ($\sigma_{\textrm{observed}}$) measured by association size metrics (see Section~\ref{sec:metrics}) is made of two components: the inherent Galactic position scatter of members of an NYA ($\sigma_{\textrm{inherent}}$) and an additional, artificial scatter due to the impact of measurement errors ($\sigma_{\textrm{error}}$). Assuming Gaussian distributions, the total observed scatter is given by:

\begin{equation}
\label{eq:scatters}
\sigma^2_{\textrm{observed}}(t) = \sigma^2_{\textrm{inherent}}(t) + \sigma^2_{\textrm{error}}(t).
\end{equation}

\noindent
 Hence, the total observed scatter (the scatter that is directly observed in the traceback analysis) is an approximation of the inherent scatter (the scatter we are trying to measure). Unlike the inherent scatter, which we assume is minimal at the epoch of stellar formation and grows over time, the scatter due to measurement errors is minimal at the current-day epoch and increases as stellar trajectories are traced back in time, due to the increasingly imprecise $XYZ$ Galactic positions that result from errors on $UVW$ space velocities, mostly on account of relatively imprecise radial velocity measurements. Adding Gaussian-like fluctuations to the current-day radial velocities of members of the core sample of \bpmg\ across all directions on the sky affects the angle of convergence of the reconstructed $UVW$ space velocities, and thus biases the epoch and $XYZ$ Galactic position where members converge. Therefore, traceback ages become not only less precise but also biased toward younger ages. The older the NYA, the longer the trajectories of its members must be traced back in time to find the epoch of minimal association size, and the greater the error and bias on traceback ages, limiting the traceback method to younger associations.

In an effort to characterize this bias on traceback ages, we constructed simulated NYA samples initialized with a range of initial $XYZ$ Galactic position and $UVW$ space velocity scatters. After initializing a set of synthetic samples following these distributions, we projected them forward in time and added a range of Gaussian-like fluctuations on the simulated radial velocity measurements and a range of radial velocity shifts (see Section~\ref{sec:sensitivity_radial_velocity_shifts}). Then, the resulting biased, noisy, reconstructed $XYZ$ Galactic positions and $UVW$ space velocities were traced back in time to determine how the epoch of minimum association size was affected. The biases (i.e., the average offsets of traceback ages from the actual age) and errors (i.e., the width of the distributions of traceback ages) on the traceback ages that result from this analysis are presented in Figure~\ref{fig:simulated_samples}.

Larger measurement errors on radial velocity result in accordingly larger biases and errors on the traceback age, due to the added, artificial $XYZ$ Galactic position scatter. To account for this bias, a correction to the traceback age is applied based on results from the simulation-derived bias that we observe for each association size metric, given that different size metrics may be more or less sensitive to this bias. These corrections are approximate, and they require assumptions on the approximate age of the association, its initial $XYZ$ Galactic position scatter, and its initial $UVW$ space velocity scatter. However, the resulting bias does not depend strongly on these assumptions as long as the initial $UVW$ space velocity scatter is not significantly smaller than $\sim 0.5$\,\kms (see Figure~\ref{fig:simulated_samples}), in which case the traceback method becomes much more sensitive to this bias. A greater $UVW$ space velocity scatter is beneficial to the precision and accuracy of traceback ages, as the point of origin becomes more clearly defined, and the divergence of stars over time is faster and more easily measurable. In contrast, the initial $XYZ$ Galactic position scatter has little impact on the bias and error on the traceback age, although a smaller initial $XYZ$ Galactic position scatter does slightly improve the error on the traceback age. This can be understood from the simple fact that the inherent $XYZ$ scatter of low-$UVW$-scatter associations would grow much more slowly over time, reducing the signal that we rely on to determine the traceback age.

In the extreme case of a null $UVW$ scatter, the NYA would not grow over time and the only change in the spatial extent of the NYA would be due to the Gaussian-like fluctuations added to the simulated current-day observations, which reduces the error on the traceback age because the bias on the traceback age is close to the actual age of the simulated NYA (i.e., the spatial extent is always minimal at the current-day epoch). In fact, with greater artificial errors on observations with a null $UVW$ space velocity scatter, the result is even more extreme: a bias on the traceback age equal to the actual age of the simulated NYA, and a null error on the traceback age. While this case is not physically realistic, it highlights how low-$UVW$-scatter NYAs are more sensitive to the bias on the traceback age due to measurement errors.

In order to compute a correction for the bias due to measurement errors in our analysis, we created simulated samples with the same $XYZ$ Galactic positions and $UVW$ space velocities, initial scatters, and measurement errors, the same number of members and the same age as those used in for simulated samples in Figure~\ref{fig:simulated_samples}, and a null radial velocity shift, as this bias is corrected independently (see Section~\ref{sec:sensitivity_radial_velocity_shifts}). With these assumptions, we computed a \effecterror\ correction to the computed traceback age of \bpmg, depending on the size metric to account for the bias due to measurement errors.

\subsubsection{Sensitivity to radial velocity shift}
\label{sec:sensitivity_radial_velocity_shifts}

Another important source of bias is the shift on radial velocity measurements due to both gravitational redshift and convective blueshift, which add a spectral-type-dependent shift of \rvshift\ (see sections~\ref{sec:grav_redshift}~and~\ref{sec:conv_blueshift}).

Our simulated samples (see Figure~\ref{fig:simulated_samples}) show that this can bias the computed traceback age of members of the core sample of \bpmg, for which the typical range of total radial velocity shift is $0.2-0.6$\,\kms, by $\sim 4.0$\,Myr. A greater bias on radial velocity increases the bias on the traceback age, although the error on the traceback age decreases slightly. Therefore, it is necessary to account for the effects of the gravitational redshift and convective blueshift on radial velocity measurements in order to derive reliable traceback ages. In order to compensate for this bias, a spectral-type-dependent correction equal to the opposite of the computed total radial velocity shift (see Figure~\ref{fig:conv_blueshift}), is applied to empirical radial velocity measurements of members of the core sample of \bpmg.

\section{Results and discussion}
\label{sec:results_discussion}

We applied the traceback method described in Section~\ref{sec:traceback_method} to the core sample of \bpmg\ members described in Section~\ref{sec:sample_selection} to derive a kinematic age estimate for the association.

\subsection{Traceback analysis of the \bpictext}
\label{sec:traceback_analysis}

\begin{figure}
 	\includegraphics{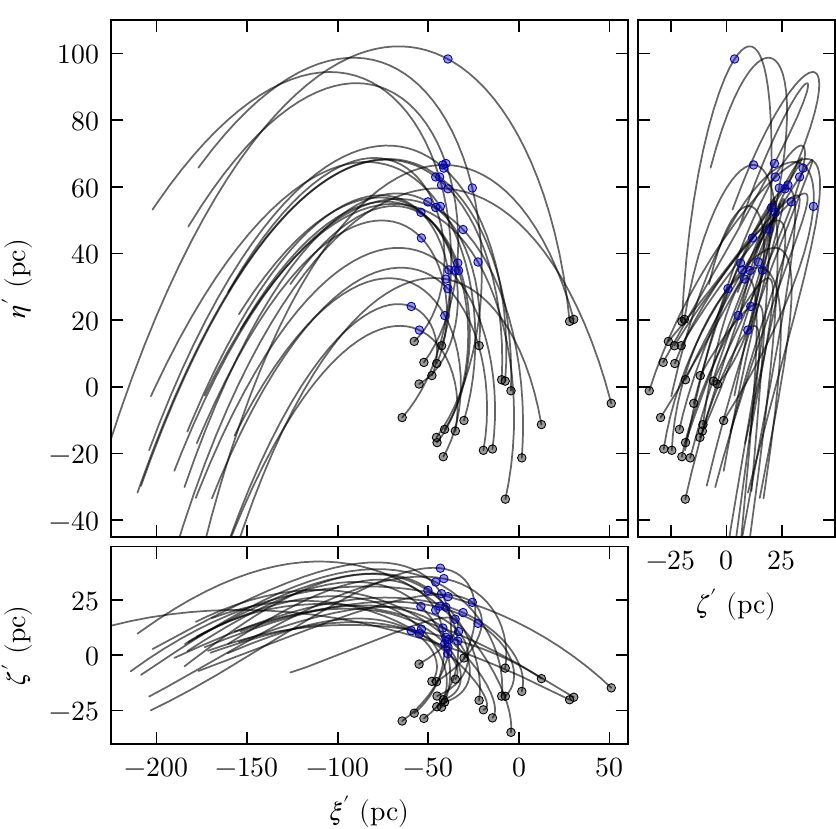}
 	\caption{Backward orbital integration of individual members of the core sample of \bpmg\ in \xez\ Galactic coordinates, back to $50$\,Myr into the past. Black circles indicate the members' positions at the current epoch, and blue circles indicate their positions at the epoch of minimal association size, as measured by the variance along the $\xi^\prime$-axis.\,This figure is adapted from Figure~3 of \cite{2020AA...642A.179M}.}
 	\label{fig:beta_pictoris_1}
\end{figure}

The traceback analysis of \bpmg, back to $50$\,Myr into the past, is presented in figures~\ref{fig:beta_pictoris_1}~and~\ref{fig:beta_pictoris_2}. With \gaiaedr\ data, one member included in the core sample, CD--31~16041, was automatically identified as a kinematic outlier (see Section~\ref{sec:kinematic_outliers}), due to its Galactic position reaching $4.3\sigma$ above the association's average Galactic position along the $\xi^\prime$-axis, beyond the $3\sigma$ threshold (se Section~\ref{sec:kinematic_outliers}). Its RUWE does not suggest it is unresolved multiple system. However, with a total of 12 available radial velocity measurements from \gaiadr\ and other kinematic surveys, this star is no longer considered an outlier and it was used in the computation of association size metrics.

As expected from their similar kinematics, members of the core sample of \bpmg\ follow similar backward Galactic orbits. In Figure~\ref{fig:beta_pictoris_2}, the dispersion along the $\xi^\prime$-axis reaches a clearly defined minimum value, which corroborates our assumption that the wider $UVW$ space velocity dispersion of members of \bpmg\ along the $U$-axis would result a greater change in association size over time. In contrast, no clear minimum is observed along the $\eta^\prime$- and $\zeta^\prime$-axes. As expected, members follow sinusoidal trajectories along the $\zeta^\prime$-axis (perpendicular to the Galactic plane) and do not reach a minimum at the epoch of stellar formation. Because members do not clearly converge across these two directions, there is little useful data along these two axes for the computation of the traceback age of \bpmg.

\subsection{Traceback age of the \bpictext}
\label{sec:traceback_age}

\begin{figure}[t]
    \includegraphics{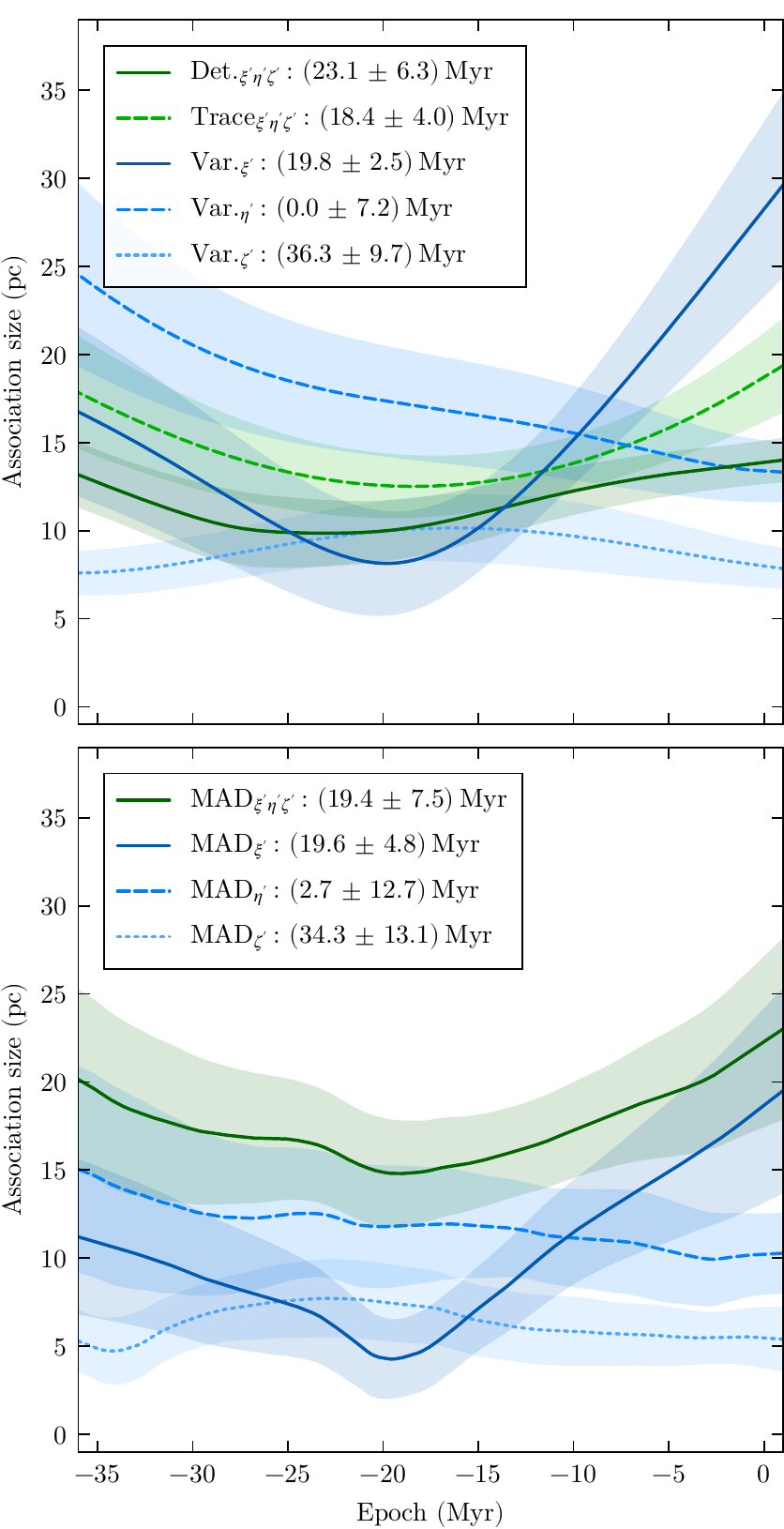}
 	\caption{\bpmg\ size as a function of time, back to 35\,Myr into the past, using only members of the core sample. Several association size metrics are used: the determinant and trace of the empirical \xez\ covariance matrix, as well as the variance along the $\xi^\prime$-, $\eta^\prime$- and $\zeta^\prime$-axes (top panel), and the total \xez\ median absolute deviation (MAD) and the MAD along the same three axes (bottom panel). The epoch of minimal association size for each metric and its associated error are indicated in the legend.}
 	\label{fig:beta_pictoris_3}
\end{figure}

Figure~\ref{fig:beta_pictoris_3} shows several association size metrics back to $35$\,Myr in the past for members of the core sample of \bpmg. The traceback ages of \bpmg\ for all association size metrics tested in this work are presented in Table~\ref{tab:size_metrics}. These results confirm that, due to the wider $UVW$ space velocity dispersion of members of \bpmg\ along the $U$-axis ($\sigma_U = 1.66$\,\kms), size metrics along the $X$- and $\xi^\prime$-axes show greater contrast (i.e., the relative change from the current-day epoch to the epoch of minimal association size), and reach a minimum value at a more distant epoch when compared to size metrics along the $Y$- ($\sigma_V = 0.52$\,\kms), $\eta^\prime$-, $Z$- ($\sigma_W = 0.73$\,\kms) or $\zeta^\prime$-axes, consistent with the result from our simulated samples (see Section~\ref{sec:simulated_samples} and  Figure~\ref{fig:simulated_samples}). Size metrics along the $Z$- and $\zeta^\prime$-axes are safely ignored, due to the lack of convergence along these axes. Compound size metrics, which include data along all axes, such as the determinant and trace of the \xez\ covariance matrix, or the total \xez\ MAD, reach a minimum value at a slightly older epoch but with an inferior contrast as a result of the narrower $UVW$ space velocity dispersion of members of the core sample of \bpmg\ along the $\eta^\prime$- and $\zeta^\prime$-axes. We also note that size metrics along the $\xi^\prime$-axis have a greater contrast than those based on the $X$-axis. As a result, we focus the remainder of our analysis on size metrics along the $\xi^\prime$-axis.

\begin{figure}
    \includegraphics{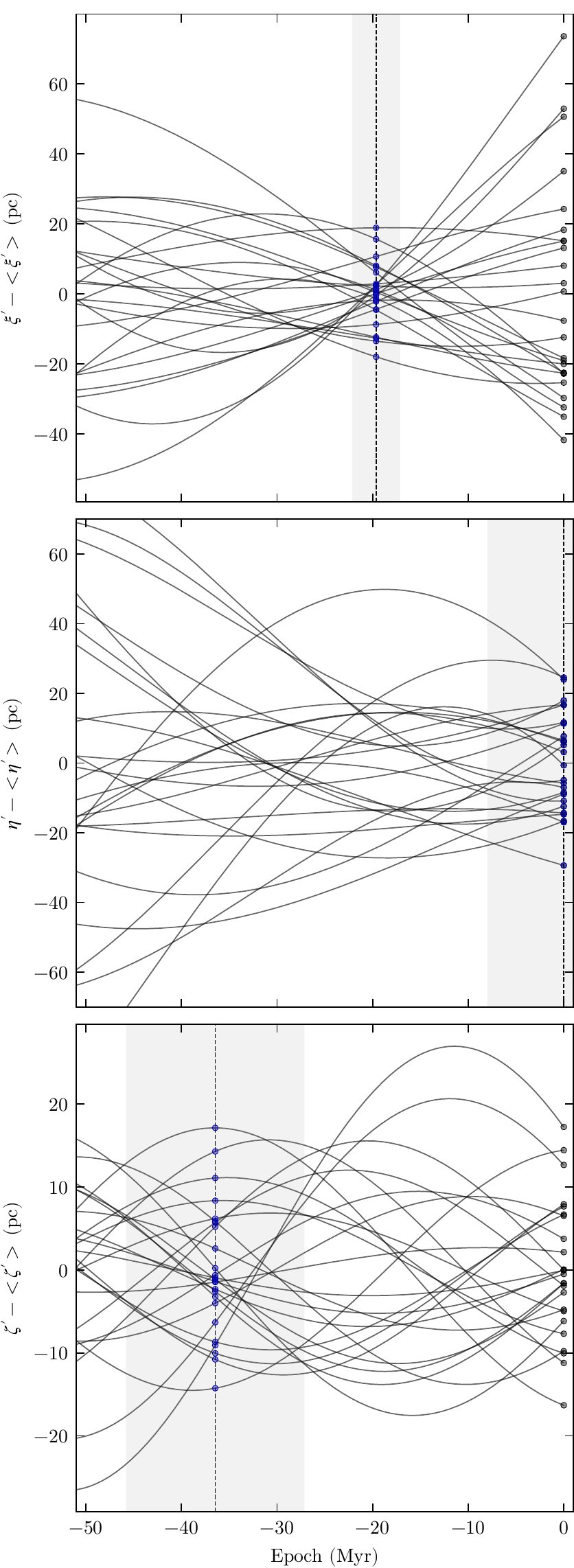}
\end{figure}

\begin{figure}[t]
    \caption{Backward orbital integration of individual members of the core sample of \bpmg\ in \xez\ Galactic coordinates as a function of time, back to $50$\,Myr into the past. The average changes in position are the result of the average orbit in the Galaxy of members of the core sample of \bpmg, and the residual changes shown in all three panels are due to individual variations in $UVW$ space velocity among members. Black circles indicate the members' positions at the current epoch, and blue circles indicate their positions at the epoch of minimal association size, as measured by the variance along the $\xi^\prime$-, $\eta^\prime$- and $\zeta^\prime$-axes. The vertical dashed line represents the epoch of smallest spatial scatter using the same three association size metrics. The $\xi^\prime$ variance provides the most signal due to the greater velocity dispersion along this axis, whereas the $\zeta^\prime$ variance provides no useful data. This is an expected result of the Galactic orbits along this axis, which follow sinusoidal paths perpendicular to the Galactic plane.}
    \label{fig:beta_pictoris_2}
\end{figure}

The contrast is slightly higher in the case of robust size metrics than equivalent empirical size metrics, but the epoch of minimal association size is similar and their jackknife error is significantly larger. This is an expected result for robust size metrics, due to the smaller number of members used by these metrics. A similar observation can be made for size metrics based on the spatial--kinematic cross-covariance matrices such as the $X-U$ and $\xi^\prime-\dot{\xi}^\prime$ cross-variances: a minimal value is reached at the same epoch as for their counterparts based on the $XYZ$ and \xez\ covariance matrices, and the contrast is the highest of any size metric tested in this study. However, the jackknife error of $\xi^\prime-\dot{\xi}^\prime$ cross-variance is much larger, suggesting that it is unsuitable for this traceback analysis.

Size metrics based on the MAD offer contrasts similar to those of their counterparts based on the spatial--kinematic cross-covariance matrices. However, their total errors are larger than those of size metrics based on the empirical $XYZ$ and \xez\ covariance matrices. As for size metrics based on the use of an MST, we observe no difference with respect to those based on the $XYZ$ or \xez\ coordinate systems. This result is not too surprising, given how MSTs are less sensitive to the shape of the NYA and only measure the change in distances between members. Nevertheless, size metrics based on the use of an MST are excluded from the current analysis, due to their low contrast.

We also tested the impact of taking into account the Galactic potential, correcting the radial velocity shifts, or minimizing the impact of the sample contamination by kinematic outliers and multiple systems on the computation of accurate traceback ages for a young NYA like \bpmg. We find traceback ages \effectorbit\ younger without considering the impact of the Galactic potential. As expected based on our discussion in Section~\ref{sec:orbit_intergration}, the impact of the Galactic potential on traceback ages is small, but not insignificant. The impact of the choice of the Galactic potential is negligible: if we use the Galactic potential from \cite{2015ApJS..216...29B} instead of model I from \cite{2013AA...549A.137I}, the final age difference is $< 0.2$\,Myr, confirming the result from \cite{2020AA...642A.179M}. If radial velocity measurements are not corrected for, the unaccounted impacts of gravitational redshift and convective blueshift cause our traceback ages to be \effectrvshift\ younger. The biggest change, however, comes from mitigating contamination by kinematic outliers and multiple systems. With the full input sample, traceback ages for \bpmg\ are \effectcontamination\ younger.

We found a raw traceback age of \ageunc\  for \bpmg\ using the $\xi^\prime$ variance as the size metric. We computed a correction to account for the bias due to measurement errors for every size metric with simulated samples representative of \bpmg\ using the same parameters as Figure~\ref{fig:simulated_samples} to account for the limited number of members and the impact of measurement errors in astrometric and kinematic data (see Section~\ref{sec:simulated_samples}). The measurement errors that were applied on simulated stars were taken from one-to-one associations between simulated stars and real members of the core sample of \bpmg. We find that the measurement errors specific to our sample alone cause a \effecterror\ bias toward younger ages, using the $\xi^\prime$ variance as the size metric. Therefore, we apply a $0.6$\,Myr correction to our raw traceback age and find a final, corrected traceback of \agecor\ for \bpmg.

\subsection{Comparison with other age-dating methods}
\label{sec:comparison_other_methods}

\begin{figure}
 	\includegraphics{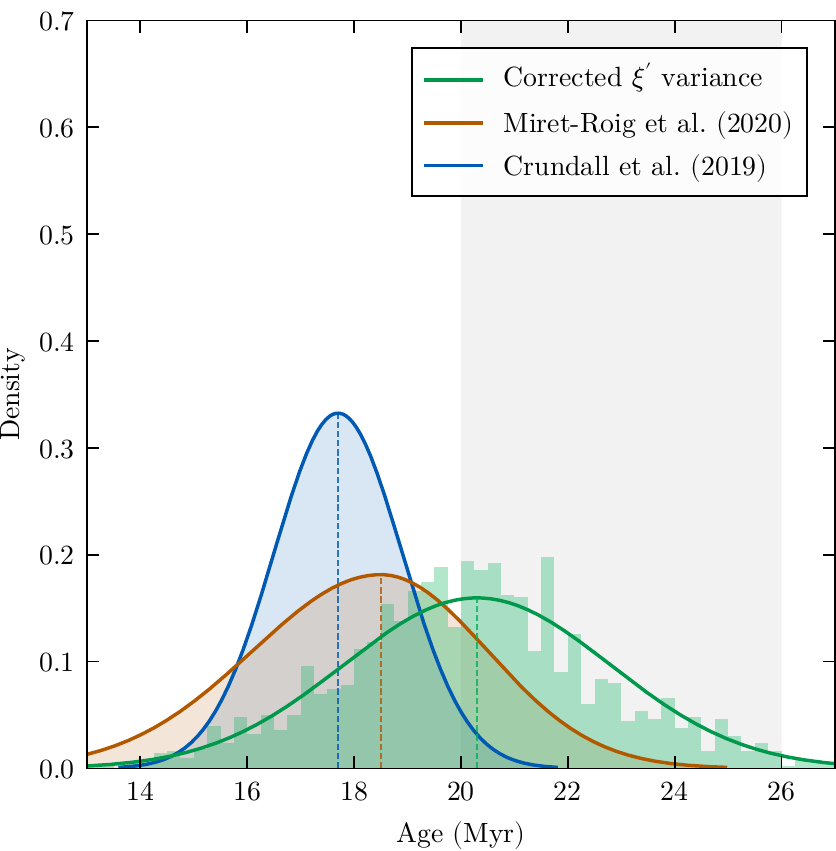}
    \caption{Traceback age distribution of \bpmg\ with 500 jackknife Monte Carlo iterations using a 50\,\% fraction, computed by minimizing the variance along the $\xi^\prime$-axis with a correction to account for the bias on traceback ages due to measurement errors (green). Our age estimate, \agecor, is compatible with other recent kinematic ages, i.e., \agemiretroigtrace\  (\citealp{2020AA...642A.179M}, orange) and \agecrundall\ (\citealp{2019MNRAS.489.3625C}, blue). However, all three age estimates remain incompatible with the range ages for \bpmg\  computed with the isochrones or LDB methods (\citealp{2014MNRAS.445.2169M}, light gray shaded area).}
 	\label{fig:beta_pictoris_4}
\end{figure}

While our corrected traceback age estimate for \bpmg\ is significantly older than previous traceback age estimates ($11-13$\,Myr; \citealp{2002ApJ...575L..75O, 2003ApJ...599..342S, 2004ApJ...609..243O, 2018AA...615A..51M}), it is compatible with several more recent studies. Figure~\ref{fig:beta_pictoris_4} shows the age distribution of \bpmg\ using the corrected $\xi^\prime$ variance as the association size metric, compared to other recent kinematic age estimates for \bpmg\ by \cite{2019MNRAS.489.3625C} and \cite{2020AA...642A.179M}. All three results are compatible with each other despite the differences in samples and methods.

The result from \cite{2019MNRAS.489.3625C}, who reported a kinematic age of \agecrundall\ for \bpmg, can be more directly compared to our results using the $X-U$ spatial--kinematic cross-covariance matrix, because this is the same metric indirectly used by the \texttt{Chronostar} method (the age of their forward model is constrained by the assumption that the initial spatial--kinematic model of \bpmg\ has no spatial--kinematic cross-covariances). Using this size metric, we report a slightly older traceback age of $19.9 \pm 3.1$\,Myr, including a $1.0$\,Myr correction to account for the biases on traceback ages due to measurement errors. This difference is mainly due to the more reliable radial velocity measurements from \gaiadr\ and other kinematic surveys used in our study.

\cite{2020AA...642A.179M} reported traceback ages of \agemiretroigtrace\ and \agemiretroigdet\ for \bpmg\ by minimizing the trace and the determinant of the \xez\ covariance matrix, respectively, computed with scikit-learn's robust estimator \citep{2011JMLR...12.2825P}. This is compatible with our result using the same association size metrics. We report corrected traceback ages of $18.6 \pm 4.0$\,Myr and $21.3 \pm 6.5$\,Myr using the same size metrics, respectively, although the errors on the traceback age are significantly larger. If instead we use the exact same sample of $26$ \emph{bona fide} members as described in \cite{2020AA...642A.179M}, along with our robust version of the trace and determinant of the \xez\ covariance matrix, we find corrected traceback ages of $21.3 \pm 8.4$\,Myr and $22.4 \pm 8.1$\,Myr, respectively. However, in our analysis, we excluded these specific size metrics in favor of size metrics with greater contrast and lower jackknife errors, such as the $\xi^\prime$ variance. The difference between our result and that of \cite{2020AA...642A.179M} is likely due not to differences in samples, but rather to the methods that were used to compute the spatial extent of the association and the Galactic orbits of its members.

Nevertheless, all of the ages measured by traceback analyses remain younger than the age estimates of \bpmg\ determined using other approaches, such as the isochrones and LDB methods ($20-26$\,Myr; \citealp{2008ApJ...689.1127M,2010ApJ...711..303Y,2014ApJ...792...37M,2014MNRAS.445.2169M,2014MNRAS.438L..11B,2015MNRAS.454..593B,2022AA...664A..70G}). While precise astrometric and kinematic measurements, clean uncontaminated samples, and corrections applied to radial velocity measurements to account for the gravitational redshift and convective blueshift helped bridge the gap with other methods, kinematic ages still remain younger than those calculated using the isochrones and LDB methods by $\sim 1-3$\,Myr.

This difference may be explained by several factors. First, there may still be contamination in the core sample of \bpmg, or a more robust way of measuring the spatial extent of the association over time might be necessary. However, the issue could also be more fundamental. Based on ages measured for star-forming regions such as Tau-Aur \citep{2021AJ....162..110K}, members may remain gravitationally bound to interstellar dust and gas for less than $\sim 2-3$\,Myr after their formation. This would bias traceback ages, regardless of the association size metric used, because the initial assumption that members follow independent Galactic orbits would only be true once the initial cloud has been dissipated and members are free of the influence of the rest of the association. In other words, the traceback method would measure the time since members of the NYA are gravitationally unbound, not the time since stellar formation. Such a systematic difference could potentially be further tested if the greater kinematic accuracy allows us to measure the traceback ages of older associations in the near future, as we may expect to find a similar offset between traceback ages and other methods.

\section{Conclusions}
\label{sec:conclusions}

In this study, we created a numerical tool capable of the deriving the age of an NYA, based solely on astrometric and kinematic measurements by tracing back the Galactic orbits of a sample of stars and evaluating the size of the association with multiple metrics. The \texttt{kanya} Python package takes into account several observational biases, including the bias on traceback ages due to measurement errors in astrometric and kinematic data, as well as the biases on absolute radial velocity measurements due to the gravitational redshift and convective blueshift of the star. Our results confirm that it is crucial to compensate for these biases and limit sample contamination by unresolved multiple systems and other kinematic outliers, in order to compute reliable traceback age estimates.

We applied our method to a core sample of \coresample\ members of \bpmg, which were assembled from the literature and include data from the \gaiadr\ catalog, and we found that minimizing the variance along the $\xi^\prime$-axis offers the least random and systematic errors, due to the wider $UVW$ space velocity dispersion of members of \bpmg\ along the $U$-axis, which tends to maximize its spatial growth along the $\xi^\prime$-axis over time. We found a corrected traceback age of \agecor, a result compatible with other recent kinematic ages found by other studies, but still slightly lower, by $\sim 1-3$\,Myr, than ages obtained using either the isochrones or LDB methods.

In future works, we plan to apply our traceback method to other known NYAs. Results will also be improved by new data from dedicated radial velocity surveys and future releases of the \gaia\ catalog, which has dramatically increased the number of stars with precise radial velocity measurements. More robust NYA modeling and association size metrics may also help further bridge the gap with ages derived using non-kinematic methods.

\section{acknowledgments}
\label{sec:acknowledgments}

We would like to thank Miret-Roig, N. and Mamajek, E. for their assistance, useful discussions, and helpful answers to our inquires. J.~G. and R.~D. acknowledge the support of the Natural Sciences and Engineering Research Council of Canada (NSERC), funding reference numbers RGPIN-2021-03121 and RGPIN-2017-06777, respectively. This work was partially carried under a Banting grant from NSERC. This work has made use of: the SIMBAD database and VizieR catalog access tool, operated at the Centre de Donn\'ees astronomiques de Strasbourg, France \citep{2000AAS..143...23O}; data products from the Two Micron All Sky Survey (\emph{2MASS}), which is a joint project of the University of Massachusetts and the Infrared Processing and Analysis Center (IPAC)/California Institute of Technology (Caltech), funded by the National Aeronautics and Space Administration (NASA), and the National Science Foundation (\citealp{2006AJ....131.1163S}; DOI: 10.26131/IRSA2); data from the European Space Agency (ESA) mission
\gaia\ (\citealp{2016AA...595A...1G}). Gaia data are being processed by the \gaia\ Data Processing and Analysis Consortium (DPAC;
\url{https://www.cosmos.esa.int/web/gaia/dpac/consortium}). Funding for the DPAC
has been provided by national institutions, in particular the institutions
participating in the \gaia\ Multilateral Agreement (MLA). The \gaia\ mission website is \url{https://www.cosmos.esa.int/gaia}. The \gaia\ archive website is \url{https://archives.esac.esa.int/gaia}. 

D.C. wrote the codes and the manuscript, and generated figures and tables; J.G. compiled the list of candidates and built figures~\ref{fig:grav_redshift}--\ref{fig:conv_blueshift}; R.D. led the analysis and provided general comments.

\software{\banyanE\ \citep{2018ApJ...856...23G}, galpy \citep{2015ApJS..216...29B}, Scikit-learn \citep{2011JMLR...12.2825P}, Extreme deconvolution \citep{2011AnApS...5.1657B}, Astropy \citep{2013AA...558A..33A, 2018AJ....156..123A, 2022ApJ...935..167A}}

\bibliographystyle{apj}
\bibliography{References}

\tabletypesize{\small}
\startlongtable
\begin{deluxetable*}{lllll}
\tablecolumns{5}
\tablecaption{Members of the input sample of \bpmg}
\label{tab:sample}
\tablehead{
    \colhead{Designation}	& \colhead{\gaiadr\ ID} & \colhead{2MASS ID}& \colhead{Spectral}            & \colhead{References$^{\textrm{b}}$}\\
    \colhead{}              & \colhead{}            & \colhead{}        & \colhead{type$^{\textrm{a}}$}	& \colhead{}}
\startdata
\sidehead{\textbf{Core Sample}}
RBS 38 & 4707563810327288192 & J00172353$-$6645124 & M2.5V & 1, 2, 3\\
GJ 2006 A & 2315841869173294080 & J00275023$-$3233060 & M3.5Ve & 2, 3\\
HD 14082 B & 107774198474602368 & J02172472$+$2844305 & G2V & 4, 5, 6\\
AG Tri A & 132362959259196032 & J02272924$+$3058246 & K8V & 4, 7, 6\\
CD$-$57 1054 & 4764027962957023104 & J05004714$-$5715255 & K8Ve & 8, 5, 9\\
V1841 Ori & 3393207610483520896 & J05004928$+$1527006 & K2IV & 10, 11\\
BD$-$21 1074 A & 2962658549474035584 & J05064991$-$2135091 & M1V & 12, 1\\
AO Men & 5266270443442455040 & J06182824$-$7202416 & K3.5V & 8, 5, 9\\
UCAC2 12510535 & 5963633872326630272 & J17024014$-$4521587 & (M2) & 13\\
HD 164249 A & 6702775135228913280 & J18030341$-$5138564 & F4.5V & 9, 5\\
UCAC4 299$-$160704 & 4050178830427649024 & J18041617$-$3018280 & (M2) & 13\\
2MASS J18092970$-$5430532 & 6653162456161626368 & J18092970$-$5430532 & (M4) & 14, 13\\
1RXS J184206.5$-$555426 & 6649788119394186112 & J18420694$-$5554254 & M3.5V & 1, 3\\
Smethells 20 & 6631685008336771072 & J18465255$-$6210366 & M1Ve & 15, 5, 9\\
CD$-$31 16041 & 6736232346363422336 & J18504448$-$3147472 & K8IVe & 15, 9\\
HD 181327 & 6643589352010758400 & J19225894$-$5432170 & F6V & 8, 5, 9\\
1RXS J192338.2$-$460631 & 6663346029775435264 & J19233820$-$4606316 & M0V & 1, 2, 16\\
UCAC4 314$-$239934 & 6754492966739292928 & J19481651$-$2720319 & (M2) & 13\\
TYC 7443$-$1102$-$1 & 6747467224874108288 & J19560438$-$3207376 & K9IVe & 1, 5, 2\\
UCAC4 284$-$205440 & 6747106443324127488 & J20013718$-$3313139 & M1V & 1, 2\\
HD 191089 & 6847146784384459648 & J20090521$-$2613265 & F5V & 1, 11\\
AU Mic & 6794047652729201024 & J20450949$-$3120266 & M0Ve & 8, 5, 16\\
GSC 06354$-$00357 & 6833292181958100224 & J21100535$-$1919573 & M2V & 1, 2\\
CPD$-$72 2713 & 6382640367603744128 & J22424896$-$7142211 & K7IVe & 15, 5, 9\\
WW PsA & 6603693881832177792 & J22445794$-$3315015 & M4IVe & 4, 9\\
\sidehead{\textbf{Extended Sample}}
HD 203 & 2337191189529102336 & J00065008$-$2306271 & F3V & 8, 5, 9\\
LP 525$-$39 & 2749470902772705408 & J00323480$+$0729271 & M3Ve & 17, 18\\
GJ 3076 & 2783378162041480832 & J01112542$+$1526214 & M5 & 1, 16\\
Barta 161 12 & 2477870708709917568 & J01351393$-$0712517 & M4.3 & 19, 3\\
PM J01538$-$1459A & 5143033331902438656 & J01535076$-$1459503 & M3V & 1, 2\\
EPIC 211046195 & 68012529415816832 & J03350208$+$2342356 & M8.5 & 19\\
StKM 1$-$433 & 66245408072670336 & J03573393$+$2445106 & M2 & 13, 20\\
51 Eri & 3205095125321700480 & J04373613$-$0228248 & F0IV & 8, 21\\
2MASS J04433761$+$0002051 & 3230008650057256960 & J04433761$+$0002051 & M9$\gamma$ & 22, 23\\
V1005 Ori & 3231945508509506176 & J04593483$+$0147007 & K8IVe & 15, 5, 9\\
LP 476$-$207 & 3291643148740384128 & J05015881$+$0958587 & M4IVe & 4, 5, 6\\
AF Lep & 3009908378049913216 & J05270477$-$1154033 & F7V & 8, 5, 11\\
UCAC4 287$-$007048 & 2901786974419551488 & J05294468$-$3239141 & M4.5 & 16, 9\\
V1311 Ori & 3216729573251961856 & J05320450$-$0305291 & M2Ve & 1, 16\\
$\beta$ Pic & 4792774797545800832 & J05471708$-$5103594 & A6V & 8, 24\\
GSC 06513$-$00291 & 2899492637251200512 & J06131330$-$2742054 & M3.5V & 1, 2, 16\\
TWA 22 A & 5355751581627180288 & J10172689$-$5354265 & M5 & 4, 25\\
$\alpha$ Cir & 5849837854861497856 & J14423039$-$6458305 & A7V & 13, 24\\
V343 Nor A & 5882581895219921024 & J15385757$-$5742273 & G8V & 8, 5, 9\\
UCAC3 89$-$246860 & 5990286858802698752 & J16120516$-$4556242 & (M4) & 13\\
TYC 8726$-$1327$-$1 & 5935776714456619008 & J16572029$-$5343316 & M3V & 1, 2, 16\\
2MASS J17020937$-$6734447 & 5814416056916139520 & J17020937$-$6734447 & (M4) & 13\\
2MASS J17092947$-$5235197 & 5924347802133154944 & J17092947$-$5235197 & (M3) & 13\\
HD 155555 A & 5811866422581688320 & J17172550$-$6657039 & G5V & 8, 5\\
CD$-$54 7336 & 5924485966955008896 & J17295506$-$5415487 & K1V & 9\\
HD 160305 & 5946515438335508864 & J17414903$-$5043279 & F9V & 26\\
UCAC4 184$-$193200 & 5921821644117633152 & J17444256$-$5315471 & (M3) & 13\\
UCAC3 74$-$428746 & 5945104588806333824 & J17483374$-$5306118 & (M2) & 13\\
UCAC3 66$-$407600 & 6648834361774839040 & J18055491$-$5704307 & (M2) & 14, 13\\
HD 165189 & 6724105656508792576 & J18064990$-$4325297 & A6V & 8, 27\\
V4046 Sgr & 4045698423617983488 & J18141047$-$3247344 & K4IVe & 15, 5, 16\\
UCAC3 63$-$445434 & 6647518766053564544 & J18161236$-$5844055 & M3.5Ve & 13, 28\\
HD 168210 & 4051081838710783232 & J18195221$-$2916327 & G3IV & 15, 5, 9\\
UCAC4 229$-$170667 & 6721666531703090432 & J18281651$-$4421477 & (M2) & 13\\
UCAC4 226$-$179372 & 6709580695582315648 & J18283524$-$4457280 & (K7) & 13\\
2MASS J18430597$-$4058047 & 6728469996119674112 & J18430597$-$4058047 & (M2) & 13\\
HD 172555 & 6438274350302427776 & J18452691$-$6452165 & A7 & 8\\
PZ Tel & 6655168686921108864 & J18530587$-$5010499 & G9IV & 8, 9\\
TYC 6872$-$1011$-$1 & 6760846563417053056 & J18580415$-$2953045 & K8IVe & 15, 5, 9\\
2MASS J19243494$-$3442392 & 6742986538895222144 & J19243494$-$3442392 & M4V & 1, 2, 3\\
UCAC3 116$-$474938 & 6747467431032539008 & J19560294$-$3207186 & M4V & 1, 2\\
SCR J2010$-$2801 & 6846651450099142016 & J20100002$-$2801410 & M2.5 & 1, 3\\
SCR J2033$-$2556 & 6800238044930953600 & J20333759$-$2556521 & M4.5V & 1, 2\\
StHA 182 & 6806301370521783552 & J20434114$-$2433534 & M3.7 & 1, 2\\
HD 199143 & 6882840883188886784 & J20554767$-$1706509 & F7V & 8, 29, 9\\
PSO J318.5338$-$22.8603 & ... & J21140802$-$2251358 & L7 & 22, 30\\
TYC 9114$-$1267$-$1 & 6400160947954197888 & J21212873$-$6655063 & K7V & 31, 2, 16\\
2E 4498 & 2694353690543010560 & J21374019$+$0137137 & M5 & 10, 14\\
HD 207043 & 6462296480342791808 & J21475527$-$5255502 & G5V & 32, 24\\
HD 213429 & 2622845684814477696 & J22311828$-$0633183 & F8V & 13, 33\\
BD$-$13 6424 & 2433191886212246784 & J23323085$-$1215513 & M0V$-$IVe & 15, 5, 9\\
\enddata
\tablecomments{(a)~Parentheses indicate that the spectral type was inferred from a photometric estimate based on the \gaiagr\ color. (b)~References are given for membership and spectral types respectively.}
\tablerefs{(1)~\citealp{2013ApJ...762...88M}; (2)~\citealp{2014ApJ...788...81M}; (3)~\citealp{2017ApJ...840...87R}; (4)~\citealp{2003ApJ...599..342S}; (5)~\citealp{1987AJ.....93..959R}; (6)~\citealp{1992AA...258..217E}; (7)~\citealp{2001ApJ...562L..87Z}; (8)~\citealp{2013ApJS..208....9P}; (9)~\citealp{2006AA...460..695T}; (10)~\citealp{2012AJ....143...80S}; (11)~\citealp{2007AJ....133.2524W}; (12)~\citealp{2009AA...508..833D}; (13)~\citealp{2018ApJ...862..138G}; (14)~\citealp{2017AJ....154...69S}; (15)~\citealp{2010AA...520A..15M}; (16)~\citealp{2019AJ....157..234S}; (17)~\citealp{2012AJ....144..109S}; (18)~\citealp{1986AJ.....92..139S}; (19)~\citealp{2012ApJ...758...56S}; (20)~\citealp{2012AJ....143..114S}; (21)~\citealp{1988mcts.book.....H}; (22)~\citealp{2014ApJ...783..121G}; (23)~\citealp{2013ApJ...772...79A}; (24)~\citealp{2006AJ....132..161G}; (25)~\citealp{1993AJ....105.1962W}; (26)~\citealp{2011MNRAS.411..117K}; (27)~\citealp{1984ApJS...55..657C}; (28)~\citealp{2018AJ....156...49R}; (29)~\citealp{1978mcts.book.....H}; (30)~\citealp{2013ApJ...777L..20L}; (31)~\citealp{2010AJ....140..119S}; (32)~\citealp{2018ApJ...860...43G}; (33)~\citealp{2003AJ....126.2048G}.}
\end{deluxetable*}
\tabletypesize{\small}
\begin{longrotatetable}
\begin{deluxetable*}{lllccccl}
\tablecolumns{8}
\tablecaption{Astrometric and kinematic measurements of members of the input  sample of \bpmg}
\label{tab:observations}
\tablehead{
    \colhead{Designation}    & \colhead{R.A.$^{\textrm{a}}$}          & \colhead{Decl.$^{\textrm{a}}$}         & \colhead{$\mu_\alpha \cos \delta$}    & \colhead{$\mu_\delta$}  & \colhead{$\pi$}  & \colhead{RV}  & \colhead{References$^{\textrm{b}}$}\\
    \colhead{}      & \colhead{(hh:mm:ss.ss)} & \colhead{(dd:mm:ss.ss)} & \colhead{(\masyr)}                      & \colhead{(\masyr)}      & \colhead{(mas)}    & \colhead{(\kms)}      & \colhead{}}
\startdata
\sidehead{\textbf{Core Sample}}
RBS 38 & 00:20:57.12 & $-$66:14:47.40 & $102.89 \pm 0.02$ & $-16.83 \pm 0.02$ & $27.16 \pm 0.02$ & $10.3 \pm 0.4$ & 1, 2, 3, 4, 5, 6\\
GJ 2006 A & 00:57:35.64 & $-$32:26:52.80 & $110.20 \pm 0.04$ & $-47.19 \pm 0.04$ & $28.55 \pm 0.04$ & $8.8 \pm 0.4$ & 1, 3, 5, 7\\
HD 14082 B & 02:21:20.88 & 28:44:40.92 & $86.86 \pm 0.03$ & $-74.17 \pm 0.03$ & $25.28 \pm 0.02$ & $4 \pm 2$ & 1, 8, 3, 9, 10, 11\\
AG Tri A & 02:52:20.28 & 30:58:23.52 & $79.54 \pm 0.02$ & $-72.13 \pm 0.01$ & $24.42 \pm 0.02$ & $6.0 \pm 0.3$ & 1, 8, 3, 12\\
CD$-$57 1054 & 05:11:48.12 & $-$57:44:35.88 & $35.39 \pm 0.01$ & $74.11 \pm 0.02$ & $37.21 \pm 0.01$ & $19.3 \pm 0.8$ & 1, 8, 2, 3, 13, 14, 15\\
V1841 Ori & 05:12:19.44 & 15:26:59.64 & $18.31 \pm 0.02$ & $-58.72 \pm 0.01$ & $18.83 \pm 0.02$ & $18 \pm 1$ & 1, 3, 16\\
BD$-$21 1074 A & 05:42:29.52 & $-$21:24:50.76 & $47.19 \pm 0.02$ & $-15.38 \pm 0.02$ & $50.43 \pm 0.02$ & $21 \pm 2$ & 1, 8, 2, 3, 10, 18\\
AO Men & 06:37:02.64 & $-$72:57:19.80 & $-7.71 \pm 0.02$ & $74.41 \pm 0.01$ & $25.57 \pm 0.01$ & $16 \pm 2$ & 1, 8, 2, 3, 13, 14, 9, 15\\
UCAC2 12510535 & 17:40:01.92 & $-$45:37:59.16 & $-20.10 \pm 0.02$ & $-137.84 \pm 0.02$ & $31.30 \pm 0.02$ & $-3.0 \pm 0.5$ & 1, 3\\
HD 164249 A & 18:45:51.12 & $-$51:21:02.16 & $2.33 \pm 0.02$ & $-86.23 \pm 0.02$ & $20.29 \pm 0.02$ & $0.2 \pm 0.4$ & 1, 8, 3, 13, 11, 15\\
UCAC4 299$-$160704 & 18:04:02.64 & $-$30:41:30.84 & $3.42 \pm 0.02$ & $-65.22 \pm 0.02$ & $18.15 \pm 0.02$ & $-7.4 \pm 0.1$ & 1, 3\\
2MASS J18092970$-$5430532 & 18:22:25.68 & $-$54:29:04.92 & $4.72 \pm 0.03$ & $-108.17 \pm 0.02$ & $25.67 \pm 0.03$ & $-1.9 \pm 0.3$ & 1, 3, 4\\
1RXS J184206.5$-$555426 & 18:31:44.76 & $-$55:05:33.00 & $12.01 \pm 0.02$ & $-79.07 \pm 0.01$ & $19.44 \pm 0.02$ & $0.6 \pm 0.2$ & 1, 3, 4, 5\\
Smethells 20 & 18:43:08.76 & $-$62:49:22.08 & $13.24 \pm 0.02$ & $-80.28 \pm 0.02$ & $19.72 \pm 0.02$ & $1.8 \pm 0.5$ & 1, 8, 2, 10, 15\\
CD$-$31 16041 & 18:41:07.44 & $-$31:12:11.52 & $17.27 \pm 0.02$ & $-72.34 \pm 0.01$ & $20.22 \pm 0.01$ & $-8 \pm 3$ & 1, 8, 19, 15\\
HD 181327 & 19:44:44.88 & $-$54:27:41.76 & $24.40 \pm 0.02$ & $-82.19 \pm 0.02$ & $20.93 \pm 0.03$ & $-0.1 \pm 0.2$ & 1, 8, 3, 13, 11, 15\\
1RXS J192338.2$-$460631 & 19:54:33.48 & $-$46:53:27.60 & $18.07 \pm 0.02$ & $-57.25 \pm 0.01$ & $14.03 \pm 0.02$ & $-0.2 \pm 0.2$ & 1, 3, 20, 5, 21, 6\\
UCAC4 314$-$239934 & 19:04:08.40 & $-$27:39:27.36 & $25.15 \pm 0.02$ & $-53.38 \pm 0.01$ & $15.47 \pm 0.02$ & $-6.26 \pm 0.05$ & 1, 3\\
TYC 7443$-$1102$-$1 & 19:01:06.24 & $-$32:52:21.36 & $33.60 \pm 0.02$ & $-68.53 \pm 0.01$ & $19.49 \pm 0.02$ & $-6.1 \pm 0.4$ & 1, 8, 2, 3, 5\\
UCAC4 284$-$205440 & 20:24:18.00 & $-$33:46:45.12 & $29.23 \pm 0.02$ & $-61.39 \pm 0.01$ & $16.68 \pm 0.02$ & $-4.0 \pm 0.6$ & 1, 8, 2, 3, 4, 10, 5\\
HD 191089 & 20:16:18.84 & $-$26:46:32.52 & $40.34 \pm 0.02$ & $-67.42 \pm 0.01$ & $19.96 \pm 0.02$ & $-7 \pm 1$ & 1, 8, 3, 13, 11, 16\\
AU Mic & 20:17:28.32 & $-$31:39:27.00 & $281.32 \pm 0.02$ & $-360.15 \pm 0.02$ & $102.94 \pm 0.02$ & $-4.5 \pm 0.6$ & 1, 8, 3, 13, 20, 11, 15\\
GSC 06354$-$00357 & 21:31:21.72 & $-$19:40:00.84 & $90.61 \pm 0.03$ & $-91.00 \pm 0.02$ & $30.90 \pm 0.03$ & $-6.2 \pm 0.8$ & 1, 2, 3, 4, 5, 21, 6\\
CPD$-$72 2713 & 22:42:18.72 & $-$71:17:38.04 & $94.85 \pm 0.01$ & $-52.38 \pm 0.01$ & $27.23 \pm 0.01$ & $7.8 \pm 0.6$ & 1, 8, 2, 3, 10, 15\\
WW PsA & 22:14:33.00 & $-$33:44:56.40 & $179.94 \pm 0.02$ & $-123.19 \pm 0.02$ & $47.92 \pm 0.03$ & $2.7 \pm 0.4$ & 1, 8, 3, 13, 5, 22, 15\\
\sidehead{\textbf{Extended Sample}}
HD 203 & 00:42:33.12 & $-$23:53:31.92 & $97.11 \pm 0.03$ & $-47.28 \pm 0.02$ & $25.16 \pm 0.03$ & $7 \pm 1$ & 1, 11, 15\\
LP 525$-$39 & 00:08:43.80 & 07:29:25.80 & $104.36 \pm 0.08$ & $-62.91 \pm 0.06$ & $28.43 \pm 0.06$ & $-3.7 \pm 0.6$ & 1, 3, 18\\
GJ 3076 & 01:51:24.48 & 15:26:20.04 & $187.6 \pm 0.8$ & $-120 \pm 1$ & $46 \pm 2$ & $2.5 \pm 0.3$ & 1, 23, 20, 5, 24, 18, 25\\
Barta 161 12 & 01:48:30.60 & $-$7:47:07.80 & $96.76 \pm 0.07$ & $-48.94 \pm 0.02$ & $26.82 \pm 0.05$ & $10 \pm 4$ & 1, 2, 26, 4, 27, 5, 7, 22\\
PM J01538$-$1459A & 01:27:44.64 & $-$14:00:08.64 & $106.72 \pm 0.04$ & $-40.60 \pm 0.03$ & $29.64 \pm 0.04$ & $9 \pm 4$ & 1, 2, 3, 5\\
EPIC 211046195 & 03:45:32.40 & 23:42:34.56 & $51.5 \pm 0.1$ & $-62.85 \pm 0.08$ & $19.72 \pm 0.09$ & $16 \pm 2$ & 1, 22\\
StKM 1$-$433 & 03:23:29.40 & 24:45:09.72 & $34.34 \pm 0.02$ & $-46.52 \pm 0.02$ & $14.55 \pm 0.02$ & $14 \pm 1$ & 1, 2, 3\\
51 Eri & 04:24:02.52 & $-$2:31:34.32 & $44.0 \pm 0.1$ & $-64.03 \pm 0.08$ & $33.44 \pm 0.08$ & $16 \pm 9$ & 1, 8, 13, 11\\
2MASS J04433761$+$0002051 & 04:54:24.84 & 00:02:03.48 & $55.3 \pm 0.2$ & $-107.3 \pm 0.2$ & $47.6 \pm 0.1$ & $17.0 \pm 0.8$ & 1, 28\\
V1005 Ori & 04:53:43.08 & 01:46:59.16 & $39.13 \pm 0.01$ & $-94.90 \pm 0.01$ & $40.99 \pm 0.01$ & $19.3 \pm 0.7$ & 1, 8, 3, 13, 12, 18\\
LP 476$-$207 & 05:29:42.36 & 09:58:57.00 & $31.66 \pm 0.04$ & $-121.57 \pm 0.02$ & $42.04 \pm 0.03$ & $18 \pm 6$ & 1, 3, 29, 11, 30, 18\\
AF Lep & 05:46:11.64 & $-$11:05:55.68 & $16.92 \pm 0.02$ & $-49.32 \pm 0.02$ & $37.25 \pm 0.02$ & $20.7 \pm 0.5$ & 1, 8, 31, 11, 16, 32, 33\\
UCAC4 287$-$007048 & 05:26:10.68 & $-$32:20:45.96 & $15.26 \pm 0.02$ & $10.77 \pm 0.02$ & $33.60 \pm 0.02$ & $22.0 \pm 0.8$ & 1, 20\\
V1311 Ori & 05:01:07.68 & $-$3:54:29.88 & $10.1 \pm 0.5$ & $-40.1 \pm 0.4$ & $29.0 \pm 0.6$ & $25 \pm 1$ & 1, 9, 3, 20, 34, 29, 10\\
$\beta$ Pic & 05:49:16.32 & $-$51:55:59.88 & $4.6 \pm 0.1$ & $83.1 \pm 0.2$ & $51.4 \pm 0.1$ & $12 \pm 7$ & 35, 1, 3, 13, 11\\
GSC 06513$-$00291 & 06:18:19.44 & $-$27:17:54.24 & $-14.2 \pm 0.2$ & $-1.2 \pm 0.2$ & $30.6 \pm 0.2$ & $25 \pm 6$ & 1, 9, 3, 20, 5\\
TWA 22 A & 10:21:38.88 & $-$53:05:33.36 & $-170.1 \pm 0.2$ & $-9.4 \pm 0.2$ & $50.5 \pm 0.2$ & $14.2 \pm 0.7$ & 1, 8, 3, 10, 36, 37\\
$\alpha$ Cir & 14:37:40.44 & $-$64:01:31.44 & $-192.5 \pm 0.1$ & $-233.5 \pm 0.1$ & $60.4 \pm 0.1$ & $7.1 \pm 0.2$ & 35, 3, 13, 14, 38, 11\\
V343 Nor A & 15:44:21.84 & $-$57:17:31.20 & $-49.86 \pm 0.06$ & $-97.87 \pm 0.07$ & $24.99 \pm 0.06$ & $3 \pm 2$ & 1, 9, 8, 3, 13, 11, 15\\
UCAC3 89$-$246860 & 16:01:16.68 & $-$45:03:34.20 & $-33.49 \pm 0.03$ & $-104.29 \pm 0.02$ & $22.91 \pm 0.03$ & $30 \pm 10$ & 1, 3\\
TYC 8726$-$1327$-$1 & 16:20:03.48 & $-$53:16:27.12 & $-12.9 \pm 0.1$ & $-85.26 \pm 0.08$ & $19.69 \pm 0.09$ & $1.4 \pm 0.1$ & 1, 3, 20, 5\\
2MASS J17020937$-$6734447 & 17:32:19.68 & $-$67:25:13.80 & $-20.36 \pm 0.02$ & $-99.53 \pm 0.02$ & $24.19 \pm 0.02$ & $5.7 \pm 0.3$ & 1, 3\\
2MASS J17092947$-$5235197 & 17:22:21.72 & $-$52:24:39.24 & $-8.83 \pm 0.06$ & $-67.94 \pm 0.05$ & $15.66 \pm 0.05$ & $1 \pm 6$ & 1, 3\\
HD 155555 A & 17:21:21.60 & $-$66:02:54.24 & $-21.44 \pm 0.01$ & $-137.17 \pm 0.01$ & $32.95 \pm 0.02$ & $-20 \pm 30$ & 1, 3, 11, 15\\
CD$-$54 7336 & 17:28:45.84 & $-$54:44:10.32 & $-5.49 \pm 0.01$ & $-63.44 \pm 0.01$ & $14.79 \pm 0.01$ & $1 \pm 1$ & 1, 8, 38, 10, 19, 15\\
HD 160305 & 17:27:15.48 & $-$50:16:31.08 & $-2.06 \pm 0.02$ & $-65.95 \pm 0.02$ & $15.20 \pm 0.02$ & $1.3 \pm 0.5$ & 1, 19\\
UCAC4 184$-$193200 & 17:10:38.64 & $-$53:44:11.40 & $-1.25 \pm 0.02$ & $-82.98 \pm 0.02$ & $18.81 \pm 0.02$ & $1.6 \pm 0.2$ & 1, 3\\
UCAC3 74$-$428746 & 17:08:26.16 & $-$53:53:47.40 & $-1.71 \pm 0.02$ & $-55.96 \pm 0.01$ & $13.05 \pm 0.02$ & $1.3 \pm 0.7$ & 1, 3\\
UCAC3 66$-$407600 & 18:28:44.04 & $-$57:55:28.20 & $0.89 \pm 0.02$ & $-72.95 \pm 0.02$ & $17.71 \pm 0.02$ & $-0.1 \pm 0.4$ & 1, 3, 4\\
HD 165189 & 18:42:28.80 & $-$43:34:27.48 & $10.9 \pm 0.1$ & $-105.84 \pm 0.08$ & $22.7 \pm 0.1$ & $-6 \pm 4$ & 1, 11\\
V4046 Sgr & 18:32:37.32 & $-$32:12:24.48 & $3.51 \pm 0.02$ & $-52.72 \pm 0.02$ & $13.99 \pm 0.02$ & $-20 \pm 20$ & 1, 3, 20, 10, 5\\
UCAC3 63$-$445434 & 18:03:06.12 & $-$58:15:52.20 & $14.14 \pm 0.02$ & $-146.54 \pm 0.02$ & $32.98 \pm 0.03$ & $3.2 \pm 0.4$ & 1, 3\\
HD 168210 & 18:58:03.36 & $-$29:43:26.40 & $4.58 \pm 0.02$ & $-46.41 \pm 0.02$ & $12.43 \pm 0.02$ & $-7 \pm 3$ & 1, 15\\
UCAC4 229$-$170667 & 18:04:07.68 & $-$44:38:11.40 & $5.01 \pm 0.02$ & $-51.28 \pm 0.02$ & $12.42 \pm 0.02$ & $-1.6 \pm 0.4$ & 1, 3\\
UCAC4 226$-$179372 & 18:08:48.84 & $-$44:02:31.20 & $4.81 \pm 0.02$ & $-49.52 \pm 0.01$ & $11.96 \pm 0.01$ & $-1.5 \pm 0.6$ & 1, 3\\
2MASS J18430597$-$4058047 & 18:46:30.00 & $-$40:01:54.12 & $12.08 \pm 0.04$ & $-71.22 \pm 0.03$ & $17.10 \pm 0.03$ & $3.9 \pm 0.2$ & 1, 3\\
HD 172555 & 18:21:44.64 & $-$64:07:41.16 & $31.95 \pm 0.09$ & $-149.7 \pm 0.1$ & $34.7 \pm 0.2$ & $-10 \pm 10$ & 1, 3, 13, 11, 15\\
PZ Tel & 18:16:28.56 & $-$50:49:08.76 & $16.27 \pm 0.02$ & $-85.52 \pm 0.02$ & $21.16 \pm 0.02$ & $-20 \pm 80$ & 1, 8, 3, 13, 11, 15\\
TYC 6872$-$1011$-$1 & 18:31:02.64 & $-$29:06:54.72 & $13.63 \pm 0.05$ & $-48.15 \pm 0.04$ & $13.45 \pm 0.04$ & $-6 \pm 2$ & 1, 8, 21, 15\\
2MASS J19243494$-$3442392 & 19:08:44.88 & $-$34:17:19.32 & $26.37 \pm 0.03$ & $-75.60 \pm 0.03$ & $19.40 \pm 0.03$ & $-4 \pm 1$ & 1, 3, 4, 5\\
UCAC3 116$-$474938 & 19:00:44.64 & $-$32:52:40.08 & $33.1 \pm 0.7$ & $-73.0 \pm 0.5$ & $19.5 \pm 0.7$ & $-10 \pm 20$ & 1, 8, 2, 3, 4, 10, 5\\
SCR J2010$-$2801 & 20:30:01.80 & $-$28:58:17.76 & $47.6 \pm 0.3$ & $-74.5 \pm 0.2$ & $21.6 \pm 0.3$ & $-7 \pm 5$ & 1, 3, 4, 5\\
\enddata
\tablecomments{(a)~All right ascensions (R.A.) and declinations (decl.) are sourced from \gaiadr\ \citep{2022arXiv220800211G}.  All positions are given at the $2016$ epoch, except for PSO J318.5338$-$22.8603, for which the epoch is $2011.563872$. (b)~References are given for proper motions ($\mu_\alpha \cos \delta$ and $\mu_\delta$), parallaxes ($\pi$) and radial velocities (RV) respectively.}
\tablerefs{(1)~\citealp{2022arXiv220800211G}; (2)~\citealp{2020AJ....160..120J}; (3)~\citealp{2020AA...642A.179M}; (4)~\citealp{2017AJ....154...69S}; (5)~\citealp{2014ApJ...788...81M}; (6)~\citealp{2013ApJ...762...88M}; (7)~\citealp{2014AJ....147..146K}; (8)~\citealp{2021AA...645A..30Z}; (9)~\citealp{2016AA...596A.116S}; (10)~\citealp{2020AA...636A..74T}; (11)~\citealp{2018AA...616A...7S}; (12)~\citealp{2006AA...460..695T}; (13)~\citealp{2007AJ....133.2524W}; (14)~\citealp{ 2007AJ....133.2524W}; (15)~\citealp{2014AA...568A..26E}; (16)~\citealp{2002AJ....123.3356G}; (17)~\citealp{2018AA...616A...1G}; (18)~\citealp{2007AN....328..889K}; (19)~\citealp{2012AJ....144....8S}; (20)~\citealp{2019AJ....157..234S}; (21)~\citealp{2013MNRAS.435.1376M}; (22)~\citealp{2012ApJ...758...56S}; (23)~\citealp{2014AJ....147...85R}; (24)~\citealp{2014AJ....147...20N}; (25)~\citealp{2001MNRAS.328...45M}; (26)~\citealp{2020AA...636A..36L}; (27)~\citealp{2014MNRAS.438L..11B}; (28)~\citealp{2016ApJS..225...10F}; (29)~\citealp{2016MNRAS.455.3345B}; (30)~\citealp{2003ApJ...599..342S}; (31)~\citealp{2013ApJS..209...33M}; (32)~\citealp{2005ApJS..159..141V}; (33)~\citealp{2001AA...379..976M}; (34)~\citealp{2018AA...618A...5D}; (35)~\citealp{2007AA...474..653V}; (36)~\citealp{2011ApJ...727....6S}; (37)~\citealp{2009AA...503..281T}; (38)~\citealp{2015AA...573A.126D}; (39)~\citealp{2004AA...418..989N}; (40)~\citealp{2016ApJ...833...96L}; (41)~\citealp{2016ApJ...819..133A}; (42)~\citealp{2018AA...614A..76J}; (43)~\citealp{2013AJ....145...41K}.}
\end{deluxetable*}
\end{longrotatetable}

\tabletypesize{\small}
\startlongtable
\begin{deluxetable*}{lcccccc}
\tablecolumns{7}
\tablecaption{Multiplicity indicators of members of the input sample of \bpmg}
\label{tab:multiple}
\tablehead{
    \colhead{Designation}	& \colhead{Number}	& \colhead{RV}	& \colhead{\gaiadr}	& \colhead{CMD}	& \colhead{WDS} & \colhead{Literature}\\
    \colhead{}		& \colhead{of RVs$^{\textrm{a}}$}    	& \colhead{Variability$^{\textrm{b}}$}	    & \colhead{RUWE$^{\textrm{1}}$}		& \colhead{Bright}	    	& \colhead{Catalog$^{\textrm{2}}$}    &\colhead{}}
\startdata
\sidehead{\textbf{Core Sample}}
RBS 38 & 29 & No & 0.950 & No & No & No\\
GJ 2006 A & 35 & No & 1.806 & No & No & No\\
HD 14082 B & 16 & No & 1.008 & No & No & No\\
AG Tri A & 27 & No & 0.892 & No & No & No\\
CD$-$57 1054 & 61 & No & 1.005 & No & No & No\\
V1841 Ori & 18 & No & 1.025 & No & No & No\\
BD$-$21 1074 A & 30 & No & 1.278 & No & No & No\\
AO Men & 38 & No & 0.843 & No & No & No\\
UCAC2 12510535 & 20 & No & 0.952 & No & No & No\\
HD 164249 A & 70 & No & 1.093 & No & No & No\\
UCAC4 299$-$160704 & 7 & No & 0.655 & No & No & No\\
2MASS J18092970$-$5430532 & 22 & No & 1.440 & No & No & No\\
1RXS J184206.5$-$555426 & 33 & No & 1.262 & No & No & No\\
Smethells 20 & 27 & No & 1.029 & No & No & No\\
CD$-$31 16041 & 12 & No & 0.961 & No & \textbf{Yes} & \textbf{Yes}\\
HD 181327 & 94 & No & 0.961 & No & No & No\\
1RXS J192338.2$-$460631 & 25 & No & 1.276 & No & No & No\\
UCAC4 314$-$239934 & 14 & No & 1.267 & No & No & No\\
TYC 7443$-$1102$-$1 & 23 & No & 1.074 & No & \textbf{Yes} & No\\
UCAC4 284$-$205440 & 25 & No & 1.091 & No & No & No\\
HD 191089 & 56 & No & 0.878 & No & No & No\\
AU Mic & 100 & No & 0.927 & No & No & No\\
GSC 06354$-$00357 & 16 & No & 1.178 & No & No & No\\
CPD$-$72 2713 & 29 & No & 0.925 & No & No & No\\
WW PsA & 34 & No & 1.139 & No & No & No\\
\sidehead{\textbf{Extended Sample}}
HD 203 & 31 & \textbf{Yes} & 0.933 & No & No & No\\
LP 525$-$39 & 7 & \textbf{Slight} & 1.159 & No & No & \textbf{Yes}\\
GJ 3076 & 12 & \textbf{Slight} & ... & No & No & \textbf{Yes}\\
Barta 161 12 & 79 & \textbf{Slight} & 1.488 & No & No & \textbf{Yes}\\
PM J01538$-$1459A & 6 & No & 1.363 & No & No & \textbf{Yes}\\
EPIC 211046195 & \textbf{1} & ... & 1.217 & No & No & No\\
StKM 1$-$433 & 27 & \textbf{Slight} & 1.341 & No & No & No\\
51 Eri & 129 & \textbf{Slight} & 1.145 & No & No & No\\
2MASS J04433761$+$0002051 & \textbf{1} & ... & 1.140 & No & No & No\\
V1005 Ori & 65 & No & 0.900 & No & \textbf{Yes} & No\\
LP 476$-$207 & 34 & \textbf{Yes} & 1.254 & No & No & \textbf{Yes}\\
AF Lep & 78 & \textbf{Yes} & 0.918 & No & No & No\\
UCAC4 287$-$007048 & 28 & ... & 1.212 & No & No & No\\
V1311 Ori & 31 & No & \textbf{33.498} & No & \textbf{Yes} & \textbf{Yes}\\
$\beta$ Pic & 1507 & \textbf{Slight} & \textbf{3.072} & No & No & \textbf{Yes}\\
GSC 06513$-$00291 & 32 & No & \textbf{12.638} & No & \textbf{Yes} & \textbf{Yes}\\
TWA 22 A & 30 & No & \textbf{8.878} & No & No & \textbf{Yes}\\
$\alpha$ Cir & 4967 & \textbf{Slight} & \textbf{3.926} & No & No & No\\
V343 Nor A & 77 & \textbf{Slight} & \textbf{8.614} & No & No & \textbf{Yes}\\
UCAC3 89$-$246860 & 16 & ... & 1.448 & No & No & No\\
TYC 8726$-$1327$-$1 & 21 & No & \textbf{4.048} & No & No & \textbf{Yes}\\
2MASS J17020937$-$6734447 & 44 & ... & 1.286 & No & No & No\\
2MASS J17092947$-$5235197 & 10 & ... & \textbf{2.803} & No & No & No\\
HD 155555 A & 3 & \textbf{Yes} & 1.061 & No & \textbf{Yes} & No\\
CD$-$54 7336 & 21 & No & 0.897 & No & \textbf{Yes} & \textbf{Yes}\\
HD 160305 & 20 & \textbf{Slight} & 0.993 & No & No & No\\
UCAC4 184$-$193200 & 19 & ... & 1.211 & No & No & No\\
UCAC3 74$-$428746 & 15 & ... & 1.067 & No & No & No\\
UCAC3 66$-$407600 & 23 & \textbf{Slight} & 1.239 & No & No & No\\
HD 165189 & 16 & No & 1.137 & No & No & \textbf{Yes}\\
V4046 Sgr & 7 & \textbf{Yes} & 0.862 & No & \textbf{Yes} & \textbf{Yes}\\
UCAC3 63$-$445434 & 17 & ... & 1.368 & No & No & No\\
HD 168210 & 3 & ... & 0.728 & \textbf{Yes} & No & No\\
UCAC4 229$-$170667 & 16 & ... & 1.082 & No & No & No\\
UCAC4 226$-$179372 & 24 & ... & 1.010 & No & No & No\\
2MASS J18430597$-$4058047 & \textbf{1} & ... & 1.727 & No & No & No\\
HD 172555 & 571 & \textbf{Yes} & 1.069 & No & No & No\\
PZ Tel & 82 & \textbf{Yes} & 0.949 & No & No & \textbf{Yes}\\
TYC 6872$-$1011$-$1 & 11 & \textbf{Slight} & \textbf{2.372} & No & No & \textbf{Yes}\\
2MASS J19243494$-$3442392 & 17 & \textbf{Slight} & 1.497 & No & No & No\\
UCAC3 116$-$474938 & 11 & \textbf{Yes} & \textbf{29.710} & No & No & \textbf{Yes}\\
SCR J2010$-$2801 & 11 & \textbf{Slight} & \textbf{8.663} & No & No & \textbf{Yes}\\
SCR J2033$-$2556 & 6 & \textbf{Slight} & 1.961 & No & No & No\\
StHA 182 & 51 & \textbf{Slight} & 1.560 & No & No & \textbf{Yes}\\
HD 199143 & 5 & \textbf{Yes} & 1.033 & No & \textbf{Yes} & \textbf{Yes}\\
PSO J318.5338$-$22.8603 & \textbf{1} & ... & ... & No & No & No\\
TYC 9114$-$1267$-$1 & 62 & \textbf{Yes} & \textbf{48.126} & No & \textbf{Yes} & \textbf{Yes}\\
2E 4498 & 7 & \textbf{Yes} & \textbf{4.981} & No & No & No\\
HD 207043 & 41 & No & 0.863 & No & No & No\\
HD 213429 & 34 & \textbf{Yes} & ... & No & No & \textbf{Yes}\\
BD$-$13 6424 & 49 & \textbf{Slight} & 1.130 & No & No & No\\
\enddata
\tablecomments{RV: radial velocity; RUWE: Renormalized Unit Weight Error; CMD: color-magnitude diagram; WDS: Washington Double Star. Bold entries indicate a positive multiple system flag. (a)~See Table~\ref{tab:observations} for references for radial velocities. (b)~RV variability is present if the standard deviation of radial velocity measurements ($\sigma_{\textrm{RV}}$) is $> 2.0$\,\kms, slight if $0.6$\,\kms$\leq \sigma_{\textrm{RV}} \leq 2.0$\,\kms, and absent if $\sigma_{\textrm{RV}} < 0.6$\,\kms.}
\tablerefs{(1)~\citealp{2022arXiv220800211G}; (2)~\citealp{2022yCat....102026M}.}
\end{deluxetable*}
\tabletypesize{\small}
\begin{longrotatetable}
\begin{deluxetable*}{llccccccl}
\tablecolumns{9}
\tablecaption{Gravitational redshift and convective blueshift on radial velocity measurements of members of the core sample of \bpmg}
\label{tab:rv_shift}
\tablehead{
    \colhead{Designation} &
    \colhead{Spectral} & 
    \colhead{Mass$^{\textrm{b}}$} &
    \colhead{Radius$^{\textrm{b}}$} &
    \colhead{$\Delta$RV${}_{\textrm{grav.}}$} &
    \colhead{$\Delta$RV${}_{\textrm{conv.}}$} &
    \colhead{$\Delta$RV${}_{\textrm{total}}$} &
    \colhead{RV${}_{\textrm{cor.}}$} &
    \colhead{References$^{\textrm{c}}$}\\
    \colhead{} &
    \colhead{Type$^{\textrm{a}}$} &
    \colhead{(\msol)} &
    \colhead{(\rsol)}&
    \colhead{(\kms)} &
    \colhead{(\kms)} &
    \colhead{(\kms)} &
    \colhead{(\kms)} &
    \colhead{}}
\startdata
RBS 38 & M2.5V & $0.40 \pm 0.04$ & $0.39 \pm 0.04$ & $0.66 \pm 0.09$ & $-0.0 \pm 0.2$ & $0.6 \pm 0.2$ & $9.6 \pm 0.6$ & 1, 2\\
GJ 2006 A & M3.5Ve & $0.26 \pm 0.03$ & $0.31 \pm 0.03$ & $0.61 \pm 0.08$ & $-0.0 \pm 0.2$ & $0.6 \pm 0.2$ & $8.2 \pm 0.6$ & 1, 2\\
HD 14082 B & G2V & $1.0 \pm 0.1$ & $1.1 \pm 0.1$ & $0.62 \pm 0.09$ & $-0.4 \pm 0.2$ & $0.3 \pm 0.2$ & $4 \pm 2$ & 1, 2\\
AG Tri A & K8V & $0.59 \pm 0.06$ & $0.60 \pm 0.05$ & $0.64 \pm 0.09$ & $-0.0 \pm 0.2$ & $0.6 \pm 0.2$ & $5.4 \pm 0.5$ & 1, 2\\
CD$-$57 1054 & K8Ve & $0.59 \pm 0.06$ & $0.60 \pm 0.05$ & $0.64 \pm 0.09$ & $-0.0 \pm 0.2$ & $0.6 \pm 0.2$ & $18.7 \pm 0.9$ & 1, 2\\
V1841 Ori & K2IV & $0.78 \pm 0.08$ & $0.77 \pm 0.07$ & $0.66 \pm 0.09$ & $-0.1 \pm 0.2$ & $0.5 \pm 0.2$ & $17 \pm 1$ & 1, 2\\
BD$-$21 1074 A & M1V & $0.49 \pm 0.05$ & $0.50 \pm 0.04$ & $0.65 \pm 0.09$ & $-0.0 \pm 0.2$ & $0.6 \pm 0.2$ & $21 \pm 2$ & 1, 2\\
AO Men & K3.5V & $0.73 \pm 0.08$ & $0.73 \pm 0.07$ & $0.65 \pm 0.09$ & $-0.1 \pm 0.2$ & $0.6 \pm 0.2$ & $16 \pm 2$ & 1, 2\\
UCAC2 12510535 & (M2) & $0.44 \pm 0.05$ & $0.43 \pm 0.04$ & $0.66 \pm 0.09$ & $-0.0 \pm 0.2$ & $0.6 \pm 0.2$ & $-3.7 \pm 0.7$ & 1, 2\\
HD 164249 A & F4.5V & $1.4 \pm 0.2$ & $1.5 \pm 0.1$ & $0.59 \pm 0.08$ & $-0.2 \pm 0.2$ & $0.4 \pm 0.2$ & $-0.2 \pm 0.6$ & 1, 2\\
UCAC4 299$-$160704 & (M2) & $0.44 \pm 0.05$ & $0.43 \pm 0.04$ & $0.66 \pm 0.09$ & $-0.0 \pm 0.2$ & $0.6 \pm 0.2$ & $-8.0 \pm 0.5$ & 1, 2\\
2MASS J18092970$-$5430532 & (M4) & $0.22 \pm 0.02$ & $0.27 \pm 0.02$ & $0.58 \pm 0.08$ & $-0.0 \pm 0.2$ & $0.6 \pm 0.2$ & $-2.5 \pm 0.6$ & 1, 2\\
1RXS J184206.5$-$555426 & M3.5V & $0.26 \pm 0.03$ & $0.31 \pm 0.03$ & $0.61 \pm 0.08$ & $-0.0 \pm 0.2$ & $0.6 \pm 0.2$ & $-0.0 \pm 0.5$ & 1, 2\\
Smethells 20 & M1Ve & $0.49 \pm 0.05$ & $0.50 \pm 0.04$ & $0.65 \pm 0.09$ & $-0.0 \pm 0.2$ & $0.6 \pm 0.2$ & $1.2 \pm 0.7$ & 1, 2\\
CD$-$31 16041 & K8IVe & $0.59 \pm 0.06$ & $0.60 \pm 0.05$ & $0.64 \pm 0.09$ & $-0.0 \pm 0.2$ & $0.6 \pm 0.2$ & $-9 \pm 3$ & 1, 2\\
HD 181327 & F6V & $1.2 \pm 0.1$ & $1.4 \pm 0.1$ & $0.59 \pm 0.08$ & $-0.3 \pm 0.2$ & $0.2 \pm 0.2$ & $-0.3 \pm 0.5$ & 1, 2\\
1RXS J192338.2$-$460631 & M0V & $0.55 \pm 0.06$ & $0.55 \pm 0.05$ & $0.65 \pm 0.09$ & $-0.0 \pm 0.2$ & $0.6 \pm 0.2$ & $-0.8 \pm 0.5$ & 1, 2\\
UCAC4 314$-$239934 & (M2) & $0.44 \pm 0.05$ & $0.43 \pm 0.04$ & $0.66 \pm 0.09$ & $-0.0 \pm 0.2$ & $0.6 \pm 0.2$ & $-6.9 \pm 0.5$ & 1, 2\\
TYC 7443$-$1102$-$1 & K9IVe & $0.56 \pm 0.06$ & $0.57 \pm 0.05$ & $0.64 \pm 0.09$ & $-0.0 \pm 0.2$ & $0.6 \pm 0.2$ & $-6.7 \pm 0.6$ & 1, 2\\
UCAC4 284$-$205440 & M1V & $0.49 \pm 0.05$ & $0.50 \pm 0.04$ & $0.65 \pm 0.09$ & $-0.0 \pm 0.2$ & $0.6 \pm 0.2$ & $-4.6 \pm 0.8$ & 1, 2\\
HD 191089 & F5V & $1.3 \pm 0.1$ & $1.4 \pm 0.1$ & $0.59 \pm 0.08$ & $-0.3 \pm 0.2$ & $0.3 \pm 0.2$ & $-8 \pm 1$ & 1, 2\\
AU Mic & M0Ve & $0.55 \pm 0.06$ & $0.55 \pm 0.05$ & $0.65 \pm 0.09$ & $-0.0 \pm 0.2$ & $0.6 \pm 0.2$ & $-5.2 \pm 0.8$ & 1, 2\\
GSC 06354$-$00357 & M2V & $0.44 \pm 0.05$ & $0.43 \pm 0.04$ & $0.66 \pm 0.09$ & $-0.0 \pm 0.2$ & $0.6 \pm 0.2$ & $-6.9 \pm 0.9$ & 1, 2\\
CPD$-$72 2713 & K7IVe & $0.63 \pm 0.07$ & $0.62 \pm 0.06$ & $0.64 \pm 0.09$ & $-0.0 \pm 0.2$ & $0.6 \pm 0.2$ & $7.2 \pm 0.8$ & 1, 2\\
WW PsA & M4IVe & $0.22 \pm 0.02$ & $0.27 \pm 0.02$ & $0.58 \pm 0.08$ & $-0.0 \pm 0.2$ & $0.6 \pm 0.2$ & $2.2 \pm 0.6$ & 1, 2\\
\enddata
\tablecomments{$\Delta$RV${}_{\textrm{grav.}}$: gravitational radial velocity shift;$\Delta$RV${}_{\textrm{conv.}}$ convective radial velocity shift; $\Delta$RV${}_{\textrm{total}}$: total radial velocity shift; RV${}_{\textrm{cor.}}$: corrected radial velocity. (a)~Parentheses indicate that the spectral type was inferred from a photometric estimate based on the \gaiagr\ color. See Table~\ref{tab:sample} for references for spectral types. (b)~See Figure~\ref{fig:masses_radii} for details on stellar masses and radii determination. (c)~References are given for stellar mass and radius, respectively.}
\tablerefs{(1)~This paper; (2)~\citealp{2013ApJS..208....9P}.}
\end{deluxetable*}
\end{longrotatetable}
\tabletypesize{\small}
\begin{longrotatetable}
\begin{deluxetable*}{lcccccc}
\tablecolumns{7}
\tablecaption{The 6D kinematics of members of the core sample of \bpmg}
\label{tab:kinematics}
\tablehead{
    \colhead{Designation}    & \colhead{$X$} & \colhead{$Y$}          & \colhead{$Z$}         & \colhead{$U$}    & \colhead{$V$}  & \colhead{$W$}\\
    \colhead{}      & \colhead{(pc)}    & \colhead{(pc)} & \colhead{(pc)} & \colhead{(\kms)}                      & \colhead{(\kms)}      & \colhead{(\kms)}}
\startdata
RBS 38 & $14.64 \pm 0.01$ & $-18.64 \pm 0.02$ & $-28.27 \pm 0.01$ & $-10.9 \pm 0.2$ & $-15.6 \pm 0.3$ & $-7.9 \pm 0.3$\\
GJ 2006 A & $4.39 \pm 0.01$ & $-1.19 \pm 0.03$ & $-34.80 \pm 0.04$ & $-10.9 \pm 0.1$ & $-16.3 \pm 0.3$ & $-9.1 \pm 0.4$\\
HD 14082 B & $-27.93 \pm 0.02$ & $19.69 \pm 0.02$ & $-20.09 \pm 0.02$ & $-11.5 \pm 0.9$ & $-17 \pm 1$ & $-8 \pm 1$\\
AG Tri A & $-30.21 \pm 0.01$ & $20.29 \pm 0.01$ & $-18.98 \pm 0.02$ & $-12.6 \pm 0.2$ & $-15.5 \pm 0.2$ & $-8.3 \pm 0.2$\\
CD$-$57 1054 & $-1.544 \pm 0.006$ & $-21.332 \pm 0.006$ & $-16.337 \pm 0.004$ & $-11.1 \pm 0.5$ & $-16.1 \pm 0.6$ & $-8.8 \pm 0.4$\\
V1841 Ori & $-50.94 \pm 0.04$ & $-5.00 \pm 0.02$ & $-14.74 \pm 0.01$ & $-13.8 \pm 0.9$ & $-16.2 \pm 0.5$ & $-9.3 \pm 0.3$\\
BD$-$21 1074 A & $-12.390 \pm 0.006$ & $-11.349 \pm 0.004$ & $-10.571 \pm 0.002$ & $-12 \pm 1$ & $-15.3 \pm 0.8$ & $-7.9 \pm 0.5$\\
AO Men & $7.573 \pm 0.008$ & $-33.69 \pm 0.01$ & $-18.542 \pm 0.007$ & $-10.6 \pm 0.8$ & $-16 \pm 1$ & $-8.4 \pm 0.7$\\
UCAC2 12510535 & $30.35 \pm 0.01$ & $-10.07 \pm 0.01$ & $-1.240 \pm 0.008$ & $-9.7 \pm 0.4$ & $-16.2 \pm 0.3$ & $-10.2 \pm 0.2$\\
HD 164249 A & $45.51 \pm 0.04$ & $-15.11 \pm 0.04$ & $-11.95 \pm 0.02$ & $-8.0 \pm 0.3$ & $-16.2 \pm 0.3$ & $-9.1 \pm 0.2$\\
UCAC4 299$-$160704 & $55.11 \pm 0.05$ & $0.83 \pm 0.03$ & $-4.04 \pm 0.02$ & $-8.5 \pm 0.2$ & $-14.6 \pm 0.1$ & $-8.49 \pm 0.08$\\
2MASS J18092970$-$5430532 & $35.10 \pm 0.03$ & $-13.21 \pm 0.03$ & $-10.82 \pm 0.02$ & $-10.9 \pm 0.2$ & $-15.0 \pm 0.2$ & $-8.0 \pm 0.1$\\
1RXS J184206.5$-$555426 & $45.21 \pm 0.04$ & $-16.65 \pm 0.04$ & $-18.43 \pm 0.02$ & $-9.0 \pm 0.2$ & $-15.4 \pm 0.2$ & $-8.1 \pm 0.1$\\
Smethells 20 & $41.79 \pm 0.03$ & $-20.87 \pm 0.03$ & $-20.09 \pm 0.02$ & $-9.8 \pm 0.3$ & $-15.2 \pm 0.4$ & $-7.7 \pm 0.2$\\
CD$-$31 16041 & $48.05 \pm 0.03$ & $3.35 \pm 0.02$ & $-11.80 \pm 0.01$ & $-10 \pm 2$ & $-15 \pm 1$ & $-8.1 \pm 0.8$\\
HD 181327 & $41.04 \pm 0.04$ & $-12.72 \pm 0.04$ & $-21.18 \pm 0.03$ & $-9.2 \pm 0.2$ & $-15.3 \pm 0.2$ & $-7.8 \pm 0.1$\\
1RXS J192338.2$-$460631 & $64.45 \pm 0.06$ & $-9.16 \pm 0.06$ & $-29.69 \pm 0.04$ & $-7.6 \pm 0.2$ & $-16.4 \pm 0.2$ & $-9.5 \pm 0.1$\\
UCAC4 314$-$239934 & $57.79 \pm 0.06$ & $13.52 \pm 0.05$ & $-26.17 \pm 0.04$ & $-8.2 \pm 0.2$ & $-15.2 \pm 0.1$ & $-8.9 \pm 0.1$\\
TYC 7443$-$1102$-$1 & $45.36 \pm 0.03$ & $6.95 \pm 0.03$ & $-23.27 \pm 0.02$ & $-9.5 \pm 0.3$ & $-15.3 \pm 0.3$ & $-8.2 \pm 0.2$\\
UCAC4 284$-$205440 & $52.45 \pm 0.04$ & $7.27 \pm 0.04$ & $-28.53 \pm 0.03$ & $-8.0 \pm 0.5$ & $-15.9 \pm 0.4$ & $-9.0 \pm 0.3$\\
HD 191089 & $42.70 \pm 0.04$ & $12.26 \pm 0.03$ & $-23.45 \pm 0.03$ & $-9 \pm 1$ & $-15.4 \pm 0.7$ & $-9.0 \pm 0.7$\\
AU Mic & $7.593 \pm 0.001$ & $1.705 \pm 0.001$ & $-5.823 \pm 0.001$ & $-10.5 \pm 0.4$ & $-16.2 \pm 0.4$ & $-9.9 \pm 0.4$\\
GSC 06354$-$00357 & $22.00 \pm 0.02$ & $12.32 \pm 0.02$ & $-20.38 \pm 0.02$ & $-10.3 \pm 0.5$ & $-15.5 \pm 0.4$ & $-9.5 \pm 0.5$\\
CPD$-$72 2713 & $19.639 \pm 0.007$ & $-19.01 \pm 0.01$ & $-24.634 \pm 0.008$ & $-10.8 \pm 0.3$ & $-15.4 \pm 0.5$ & $-7.4 \pm 0.3$\\
WW PsA & $9.542 \pm 0.005$ & $2.117 \pm 0.008$ & $-18.46 \pm 0.01$ & $-10.7 \pm 0.2$ & $-16.2 \pm 0.3$ & $-9.8 \pm 0.3$\\
\enddata
\end{deluxetable*}
\end{longrotatetable}
\tabletypesize{\small}
\startlongtable
\begin{deluxetable*}{lccccc}
\tablecolumns{6}
\tablecaption{Traceback Ages of \bpmg}
\label{tab:size_metrics}
\tablehead{
    \colhead{Association Size Metric}	& \colhead{Age}	    & \colhead{Jackknife Error$^{\textrm{a}}$}	& \colhead{Measurement Error$^{\textrm{a}}$}	& \colhead{Total Error$^{\textrm{a}}$}	& \colhead{Contrast$^{\textrm{b}}$}\\
    \colhead{}			    & \colhead{(Myr)}	& \colhead{(Myr)}			    & \colhead{(Myr)}				& \colhead{(Myr)}		& \colhead{(\%)}}
\startdata
\sidehead{\textbf{Spatial Covariance Matrix}}
\bm{$X$} \textbf{Variance} & \bm{$19.0$} & \bm{$2.3$} & \bm{$1.6$} & \bm{$3.1$} & \bm{$-57.4$}\\
$Y$ Variance & $1.3$ & $8.7$ & $4.2$ & $7.6$ & $-0.1$\\
$Z$ Variance & $36.8$ & $12.6$ & $6.5$ & $14.0$ & $-4.3$\\
Determinant${{}}_{{XYZ}}$ & $21.0$ & $4.2$ & $3.7$ & $6.5$ & $-60.6$\\
Trace${{}}_{{XYZ}}$ & $18.2$ & $3.2$ & $1.9$ & $4.1$ & $-34.2$\\
$X$ Variance (robust) & $20.0$ & $3.4$ & $2.3$ & $4.1$ & $-63.4$\\
$Y$ Variance (robust) & $30.7$ & $14.7$ & $8.9$ & $12.1$ & $-5.2$\\
$Z$ Variance (robust) & $40.0$ & $12.2$ & $1.8$ & $11.8$ & $-18.3$\\
Determinant${{}}_{{XYZ}}$ (robust) & $21.3$ & $5.0$ & $3.4$ & $6.0$ & $-75.4$\\
Trace${{}}_{{XYZ}}$ (robust) & $20.8$ & $6.8$ & $3.5$ & $6.4$ & $-35.3$\\
\bm{$\xi^\prime$} \textbf{Variance} & \bm{$19.8$} & \bm{$1.5$} & \bm{$1.5$} & \bm{$2.5$} & \bm{$-71.1$}\\
$\eta^\prime$ Variance & $0.0$ & $6.2$ & $3.5$ & $7.2$ & $0.0$\\
$\zeta^\prime$ Variance & $36.3$ & $7.5$ & $2.0$ & $9.7$ & $-4.3$\\
Determinant${{}}_{{\xi^\prime \eta^\prime \zeta^\prime}}$ & $23.1$ & $3.8$ & $3.8$ & $6.3$ & $-60.5$\\
Trace${{}}_{{\xi^\prime \eta^\prime \zeta^\prime}}$ & $18.4$ & $3.2$ & $1.9$ & $4.0$ & $-34.0$\\
$\xi^\prime$ Variance (robust) & $20.0$ & $3.7$ & $2.1$ & $4.1$ & $-72.3$\\
$\eta^\prime$ Variance (robust) & $0.0$ & $12.5$ & $9.5$ & $11.3$ & $0.0$\\
$\zeta^\prime$ Variance (robust) & $40.0$ & $12.2$ & $1.8$ & $11.8$ & $-18.3$\\
Determinant${{}}_{{\xi^\prime \eta^\prime \zeta^\prime}}$ (robust) & $21.3$ & $4.9$ & $3.4$ & $6.0$ & $-75.3$\\
Trace${{}}_{{\xi^\prime \eta^\prime \zeta^\prime}}$ (robust) & $20.8$ & $6.8$ & $3.5$ & $6.3$ & $-35.1$\\
\sidehead{\textbf{Spatial--Kinematic Cross-Covariance Matrix}}
$X-U$ Cross-Covariance & $18.9$ & $2.3$ & $1.7$ & $3.1$ & $-75.8$\\
$Y-V$ Cross-Covariance & $0.0$ & $11.1$ & $4.9$ & $8.7$ & $0.0$\\
$Z-W$ Cross-Covariance & $35.7$ & $11.6$ & $8.7$ & $11.5$ & $-31.7$\\
Determinant${{}}_{{XUYVZW}}$ & $19.7$ & $8.1$ & $2.9$ & $8.4$ & $-63.5$\\
Trace${{}}_{{XUYVZW}}$ & $18.7$ & $3.2$ & $1.9$ & $4.1$ & $-74.0$\\
$X-U$ Cross-Covariance (robust) & $20.5$ & $2.9$ & $2.4$ & $4.0$ & $-65.8$\\
$Y-V$ Cross-Covariance (robust) & $34.2$ & $11.8$ & $9.6$ & $11.0$ & $-23.8$\\
$Z-W$ Cross-Covariance (robust) & $40.0$ & $11.9$ & $4.7$ & $11.8$ & $-13.8$\\
Determinant${{}}_{{XUYVZW}}$ (robust) & $25.1$ & $8.7$ & $2.9$ & $8.4$ & $-70.2$\\
Trace${{}}_{{XUYVZW}}$ (robust) & $15.7$ & $5.1$ & $2.7$ & $4.9$ & $-62.3$\\
$\xi^\prime-\dot{\xi}^\prime$ Cross-Covariance & $19.5$ & $1.7$ & $1.7$ & $2.9$ & $-83.4$\\
$\eta^\prime-\dot{\eta}^\prime$ Cross-Covariance & $0.2$ & $7.9$ & $2.7$ & $8.4$ & $-0.0$\\
$\zeta^\prime-\dot{\zeta}^\prime$ Cross-Covariance & $35.7$ & $11.6$ & $8.7$ & $11.5$ & $-31.7$\\
Determinant${{}}_{{\xi^\prime \dot{\xi}^\prime \eta^\prime \dot{\eta}^\prime \zeta^\prime \dot{\zeta}^\prime}}$ & $18.7$ & $8.4$ & $3.6$ & $8.6$ & $-48.4$\\
Trace${{}}_{{\xi^\prime \dot{\xi}^\prime \eta^\prime \dot{\eta}^\prime \zeta^\prime \dot{\zeta}^\prime}}$ & $20.5$ & $5.8$ & $2.2$ & $7.3$ & $-74.8$\\
$\xi^\prime-\dot{\xi}^\prime$ Cross-Covariance (robust) & $20.0$ & $6.4$ & $2.8$ & $4.9$ & $-75.6$\\
$\eta^\prime-\dot{\eta}^\prime$ Cross-Covariance (robust) & $6.6$ & $7.5$ & $3.1$ & $8.6$ & $-8.0$\\
$\zeta^\prime-\dot{\zeta}^\prime$ Cross-Covariance (robust) & $40.0$ & $12.0$ & $4.7$ & $11.8$ & $-13.8$\\
Determinant${{}}_{{\xi^\prime \dot{\xi}^\prime \eta^\prime \dot{\eta}^\prime \zeta^\prime \dot{\zeta}^\prime}}$ (robust) & $25.4$ & $8.2$ & $4.0$ & $9.2$ & $-60.4$\\
Trace${{}}_{{\xi^\prime \dot{\xi}^\prime \eta^\prime \dot{\eta}^\prime \zeta^\prime \dot{\zeta}^\prime}}$ (robust) & $29.0$ & $6.5$ & $6.3$ & $7.7$ & $-65.1$\\
\sidehead{\textbf{Median Absolute Deviation}}
MAD${{}}_{{X}}$ & $20.8$ & $5.3$ & $2.6$ & $5.5$ & $-64.3$\\
MAD${{}}_{{Y}}$ & $18.9$ & $10.8$ & $8.0$ & $10.7$ & $-2.8$\\
MAD${{}}_{{Z}}$ & $34.4$ & $13.6$ & $9.4$ & $13.7$ & $-10.8$\\
MAD${{}}_{{XYZ}}$ & $19.5$ & $7.1$ & $3.0$ & $7.0$ & $-36.0$\\
MAD${{}}_{{\xi^\prime}}$ & $19.6$ & $3.8$ & $1.5$ & $4.8$ & $-77.2$\\
MAD${{}}_{{\eta^\prime}}$ & $2.7$ & $11.7$ & $7.4$ & $12.7$ & $-3.3$\\
MAD${{}}_{{\zeta^\prime}}$ & $34.3$ & $13.6$ & $9.4$ & $13.1$ & $-10.8$\\
MAD${{}}_{{\xi^\prime \eta^\prime \zeta^\prime}}$ & $19.4$ & $7.1$ & $2.2$ & $7.5$ & $-34.7$\\
\sidehead{\textbf{Minimum Spanning Tree}}
MST${{}}_{{XYZ}}$ Mean & $24.3$ & $3.0$ & $2.9$ & $3.5$ & $-33.9$\\
MST${{}}_{{XYZ}}$ Mean (robust) & $18.7$ & $0.9$ & $2.9$ & $3.4$ & $-36.9$\\
MST${{}}_{{XYZ}}$ MAD & $22.6$ & $4.0$ & $5.7$ & $7.3$ & $-57.9$\\
MST${{}}_{{\xi^\prime \eta^\prime \zeta^\prime}}$ Mean & $24.3$ & $3.0$ & $2.9$ & $3.5$ & $-33.8$\\
MST${{}}_{{\xi^\prime \eta^\prime \zeta^\prime}}$ Mean (robust) & $18.7$ & $0.8$ & $2.9$ & $3.3$ & $-36.8$\\
MST${{}}_{{\xi^\prime \eta^\prime \zeta^\prime}}$ MAD & $22.6$ & $4.0$ & $5.7$ & $7.3$ & $-57.8$\\
\enddata
\tablecomments{Association size metrics are split across four categories based on the way they are computed: using the spatial covariance matrix, the spatial--kinematic cross-covariance matrix, the median absolute deviation (MAD), or the minimum spanning tree (MST). Both the $XYZ$ and the \xez\ coordinate systems are considered, as well as both empirical and robust covariance matrices metrics. Lines in bold characters are indicative of our most reliable size metrics. No correction to account for the effects of measurement errors was applied (see Section~\ref{sec:sensitivity_kinematic_errors}) (a)~The jackknife error is the error on the traceback age due to the sensitivity to sample definition, the measurement error is the error on the traceback age due to errors on astrometric and kinematic data, and the total error is computed with a Monte Carlo approach including both the jackknife and measurement errors (see Section~\ref{sec:error_traceback_age}). (b)~The contrast represent the change in association size at the epoch of minimal association size relative to the current-day epoch.}
\end{deluxetable*}
\hphantom{1}

\end{document}